\documentclass[superscriptaddress,aps,journal=prl,preprint]{revtex4-2}
\usepackage{graphicx}
\usepackage{epstopdf}
\usepackage{bm,amsmath,amssymb} 
\usepackage{color, soul} 
\usepackage{dcolumn}   
\usepackage{multirow}
\usepackage{hyperref}
\newcommand{\etal}{{\em et al}.\ }
\newcommand{\Von}{V$_{\rm O}^{0}$}
\newcommand{\Vop}{V$_{\rm O}^{2+}$}

\begin{document}
\title{Structural Polymorphism Kinetics Promoted by Charged Oxygen Vacancies in HfO$_2$}
\author{Li-Yang Ma}
\affiliation{Fudan University, Shanghai 200433, China}
\affiliation{Key Laboratory for Quantum Materials of Zhejiang Province, Department of Physics, School of Science, Westlake University, Hangzhou, Zhejiang 310024, China}
\author{Shi Liu}
\email{liushi@westlake.edu.cn}
\affiliation{Key Laboratory for Quantum Materials of Zhejiang Province, Department of Physics, School of Science, Westlake University, Hangzhou, Zhejiang 310024, China}
\affiliation{Institute of Natural Sciences, Westlake Institute for Advanced Study, Hangzhou, Zhejiang 310024, China}

\date{\today}

\begin{abstract}
Defects such as oxygen vacancy are widely considered to be critical for the performance of ferroelectric HfO$_2$-based devices, and yet atomistic mechanisms underlying various exotic effects such as wake-up and fluid imprint remain elusive. Here, guided by a lattice-mode-matching criterion,  we systematically study the phase transitions between different polymorphs of hafnia under the influences of neutral and positively charged oxygen vacancies using a first-principles-based variable-cell nudged elastic band technique.
We find that the positively charged oxygen vacancy can promote the transition of various nonpolar phases to the polar phase kinetically, enabled by a transient high-energy tetragonal phase and extreme charge-carrier-inert ferroelectricity of the polar $Pca2_1$ phase. 
The intricate coupling between structural polymorphism kinetics and the charge state of the oxygen vacancy has important implications for the origin of ferroelectricity in HfO$_2$-based thin films as well as wake-up, fluid imprint, and inertial switching.

\end{abstract}
\maketitle
\newpage
Ferroelectric memory has long been considered as a competitive non-volatile information storage technology because of various merits such as fast switching rate, low power consumption, and high endurance~\cite{Scott07p954,Huang18p1700560,Mikolajick20p1434}. Conventional perovskite ferroelectrics are not compatible with the modern complementary metal oxide semiconductor (CMOS) technology, severely limiting the scalability~\cite{Pinnow04pK13}. The discovery of ferroelectricity in silicon-doped thin films of hafnium oxide in 2011~\cite{Boscke11p102903} quickly made HfO$_2$ a leading candidate material
for incorporating ferroelectric functionalities into integrated circuits. Thanks to its robust electric dipoles at the nanoscale~\cite{Lee20p1343}, simple composition, and the CMOS compatibility, HfO$_2$-based ferroelectrics have already been implemented in several memory devices~\cite{Luo20p1391,Kim21peabe1341}.

The commercialization of ferroelectric HfO$_2$-based devices is mainly hindered by the device reliability issues associated with the profound wake-up effect, strong imprint, and limited endurance~\cite{Mulaosmanovic21p9405215,Mueller13p93}. For example, the imprint effect, manifested as the shift of a hysteresis loop along the voltage axis with time, will destabilize one of the polarization states and cause retention loss~\cite{Buragohain19p35115}. It is widely accepted that the polar orthorhombic (PO) $Pca2_1$ phase of HfO$_2$ is responsible for the ferroelectricity~\cite{Park15p1811,Huan14p064111,Sebastian14p140103,Sang15p162905,Materlik15p134109}, despite being higher in energy than the nonpolar monoclinic (M) $P2_1/c$ phase. The stabilization of the PO phase in thin films has been attributed to the concerted effects of extrinsic factors such as surface/interface effects of grains~\cite{Park15p1811,Polakowski15p232905,Materlik15p134109,Batra16p172902,Knneth17p205304,Park17p9973}, clamping stress~\cite{Shiraishi16p262904,Batra17p4139,Liu19p054404}, defects~\cite{Schroeder14p08LE02,Starschich17p333,Park17p4677,Xu17p124104,Batra17p9102,Xu16p091501,Pal17p022903}, and electric fields~\cite{Batra17p4139}. The as-prepared  HfO$_2$-based thin film often contains high volume factions of nonpolar M phase and tetragonal (T) $P4_2/nmc$ phase~\cite{Grimley18p1701258,Park17p4677,Richter17p1700131}. The models explaining the performance instability during electrical cycling can be categorized into two groups: (i) defects and their evolution; (2) electric field-induced structural polymorphism that may originate from defects~\cite{Chouprik21p11635}.
In this work, we discover an intricate coupling between structural polymorphism kinetics and the charge state of the oxygen vacancy in HfO$_2$ that has important implications for the origin of ferroelectricity and several device reliability issues. \\

\begin{figure}
	\begin{center}
		\includegraphics[width=6in]{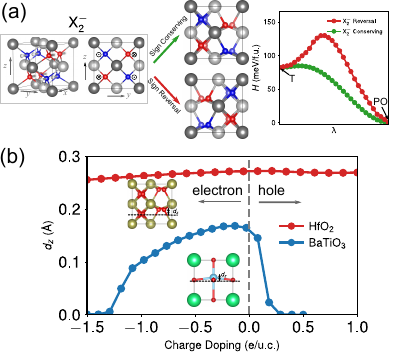}
	\end{center}
	\caption{(a) Schematic illustration of X$_2^-$ mode in the tetragonal phase of HfO$_2$. The tetragonal unit cell has the long axis along $x$. The closer and further Hf atoms are colored in dark and light grey; outward and inward displaced oxygen atoms are colored in blue and red, respectively. The generic phase space coordinate is denoted as $\lambda$. (b) Polar atomic displacement ($d_z$) as a function of charge-carrier concentration for the PO phase of HfO$_2$ and tetragonal BaTiO$_3$. The insets illustrate the definitions of $d_z$ in HfO$_2$ and BaTiO$_3$. \label{features}}
\end{figure}

We start by emphasizing two unique features of the ferroelectric PO phase of HfO$_2$. First, the X$_2^-$ lattice mode characterized by antiparallel $x$-displacements of neighboring oxygen atoms (Fig.~\ref{features}a) plays an important role in structural polymorphism of HfO$_2$.
Though phase transitions between different polymorphs of HfO$_2$ have been investigated with density functional theory (DFT) calculations previously~\cite{Liu19p054404}, the importance of the X$_2^-$ mode matching was not appreciated. There exist multiple pathways connecting two phases depending on the choice of atom-to-atom mapping. We find that an atomic mapping that conserves the sign of the X$_2^-$ mode gives a lower enthalpy barrier than a mapping that reverses the X$_2^-$ mode. As shown in Fig.~\ref{features}a, the X$_2^-$-sign-conserving pathway of  T$\rightarrow$PO has an enthalpy barrier of 2.5~meV per formula unit (f.u.), much lower than the pathway with the X$_2^-$ mode reversed (47.5~meV/f.u.). Recent DFT studies also found that the polarization switching in PO that reverses the X$_2^-$ mode needs to overcome a large barrier~\cite{Qi21parXiv}. The small enthalpy barrier of T$\rightarrow$PO already hints at the importance of nonopolar-polar structural polymorphism in HfO$_2$-based devices. 

Another intriguing feature is the extreme charge-carrier-inert ferroelectricity of the PO phase. We compute the polar atomic displacements in ferroelectric tetragonal BaTiO$_3$ and the PO phase of HfO$_2$ as a function of charge-carrier concentration obtained via the background-charge method~\cite{VASPtheguide}. Both hole doping and electron doping quickly suppress the polar displacements in tetragonal BaTiO$_3$~\cite{Michel21p8640}. In contrast, the local atomic displacements of polar oxygen atoms in PO are insensitive to charge-carrier doping (Fig.~\ref{features}b). Such feature, likely resulting from the improper nature of ferroelectric HfO$_2$~\cite{Delodovici21p064405}, suggests that the PO phase can support a substantial amount of charged defects while maintaining the polar structure. Since the charge injection to and entrapment from defects occur frequently during the electrical cycling, understanding the impacts of charged defects on polymorphism kinetics is expectedly important.

Here, we focus on neutral oxygen vacancy (V$_{\rm O}^0$) and doubly positively charged oxygen vacancy (V$_{\rm O}^{2+}$), as they are known prominent defects in HfO$_2$-based thin films~\cite{Chouprik21p11635,Guha07p196101}, particularly at the electrode/HfO$_2$ interface~\cite{Cho08p233118,Hamouda20p252903,Fengler18p1700547}. 
Different from several previous DFT studies on oxygen-deficient HfO$_2$ where the assumed vacancy concentration reaches a rather high level (up to 12.5~f.u.\%~\cite{Rushchanskii21p087602}), we focus on a relatively low vacancy concentration (3.125 f.u.\%).
DFT calculations are performed using Vienna Ab initio Simulation Package (\texttt{VASP}) with Perdew-Burke-Ernzerhof (PBE) density functional~\cite{Perdew96p3865}. 
Structural parameters of unit cells of hafnia polymorphs are optimized using a plane-wave cutoff of 600~eV, a $4 \times 4 \times 4$ Monkhorst-Pack $k$-point grid for Brillouin zone sampling, an energy convergence threshold of $10^{-6}$ eV, and a force convergence threshold of $10^{-3}$ eV/\AA. All possible configurations of a 48-atom supercell with an oxygen vacancy concentration of 3.125 f.u.\%~are explored to identify the lowest-energy configuration. 
The minimum energy pathway (MEP) connecting two polymorphs is identified with the variable-cell nudged elastic band (VCNEB) technique implemented in the \texttt{USPEX} code~\cite{Oganov06p244704,Lyakhov13p1172,Oganov11p227}. The main difference between conventional NEB and VCNEB is that VCNEB allows lattice constants to change during solid-solid transformations, thus capable of quantifying the intrinsic transition barrier between two phases possessing different lattice constants. It is noted that lattice constants and phase transition barriers predicted with PBE are comparable with values computed with SCAN~\cite{Sun15p036402}. More computational details are reported in Supplementary Material (SM)\cite{SI} that includes Refs.~\cite{MacLaren09pG103,Hoffmann15p072006,Wu20p101482,Liu19p054404,Fang13p92,Quinteros14p024501,Duncan16p400,Hou16p123106,Singh18p014501,Antad21p54133,Clima12p133102,Kamiya13p155201,Dirkmann18p14857,Kwon10p148,Kang22p731,Zhang19p1800833,Grimley16p1600173,Mittmann20p18.4.1,Buragohain19p35115,Fan16p012906,He21pL180102,Nukala21p630}.

\begin{figure}
	\begin{center}
		\includegraphics[width=6in]{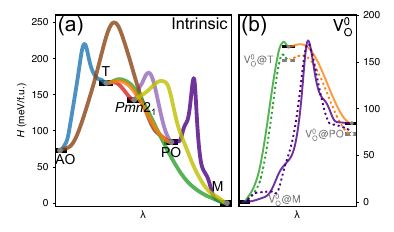}
	\end{center}
	\caption{(a) Minimum energy pathways connecting different polymorphs of HfO$_2$ obtained with VCNEB. (b) Comparison of transition pathways between M, T, and PO phases at the pristine state (solid lines) and those in the presence of V$_{\rm O}^0$ at a concentration of 3.125~\% (dashed lines). The energy of the M phase is chosen as the energy zero point. \label{barriers}}
\end{figure}

We first examine the intrinsic transition barriers in the absence of defects between different phases of HfO$_2$ including an ``antiferroelectric-like" orthorhombic (AO) $Pbca$ phase and another polar orthorhombic $Pmn2_1$ phase that may form in (111)-oriented Hf$_{0.5}$Zr$_{0.5}$O$_2$ thin films~\cite{Wei18p1095,Qi20p257603}. All pathways conserve the sign of the X$_2^-$ mode (Fig.~S1 in SM\cite{SI}). Taking the lowest-energy M phase as the reference, the energy of T, AO, PO, and $Pmn2_1$ phase is 166, 73, 84, and 143 meV/f.u., respectively. The transitions of T$\rightarrow$M, T$\rightarrow$PO, and T$\rightarrow Pmn2_1$ are all kinetically fast with negligible enthalpy barriers ($<5$~meV/f.u., \ref{barriers}a), indicating the precursor role of the high-temperature T phase~\cite{Liu19p054404}.  The transition of T$\rightarrow$AO needs to overcome a large kinetic barrier despite AO being the second most favored phase thermodynamically.  
The transitions from the polar $Pmn2_1$ phase to M and $Pca2_1$ only needs to overcome a small barrier of 28 and 37 meV/f.u., respectively. We suggest that although the emergence of the $Pmn2_1$ phase from the T phase is effortless kinetically, it may quickly evolve to other phases such that its transient presence during crystallization is probably undetectable in experiments.

Phase transitions of $Pca2_1$ to other nonpolar phases are all hindered by large barriers. Overall, the formation of the M phase is the most favored process both kinetically and thermodynamically in the absence of extrinsic effects. This explains why the most stable M phase dominates in bulk synthesis.  
We discover that  M$\rightarrow$PO involves another lattice mode, X$_5^y$. It is impossible for the transition pathway to conserve both X$_5^y$ and X$_2^-$, thus making M$\rightarrow$PO a high-barrier (169 meV/f.u.) process that can not be activated by an electric field alone (see Sec.~\uppercase\expandafter{\romannumeral1}.C in SM\cite{SI}).
This is inconsistent with experimental observations that a substantial amount of M phase is transformed to the PO phase in the woken-up thin films of HfO$_2$~\cite{Sang15p162905,Hoffmann15p072006,Grimley16p1600173}. As we will discuss in below, the presence of charged oxygen vacancies resolves this puzzle.    

The oxygen vacancy has long been postulated to be important for the stabilization of the PO phase. Experimentally, a low oxygen content during the deposition favors the formation of T and PO phases whereas a high oxygen content tends to stabilize the M phase~\cite{Baumgarten21p032903,Mittmann19p1900042}. However, DFT calculations suggested that even at a high V$_{\rm O}^0$ concentration of $\approx$12.5 f.u.\%, the M phase remains much more stable thermodynamically than the PO phase~\cite{Zhou19p143}.
Rushchanskii~\etal~proposed that configurations with two-dimensional extended oxygen vacancies tend to transform into the PO phase~\cite{Rushchanskii21p087602}, but the V$_{\rm O}^0$ concentration in their model reached 12.5 f.u.\%, much higher than typical experimental values of 2--3 f.u.\%~\cite{Mittmann19p1900042,Pal17p022903}. We further investigate the effects of V$_{\rm O}^0$ on polymorphism kinetics at a concentration of 3.125 f.u.\%.
Compared with intrinsic defect-free cases, the presence of V$_{\rm O}^0$ has little impact on the relative stability, nor does it substantially affect the kinetics of phase transitions (Fig.~\ref{barriers}b). We conclude that at experimentally relevant concentrations, V$_{\rm O}^0$ alone cannot explain the strong impacts of O$_2$ partial pressures on the ferroelectric properties of HfO$_2$-based thin films.

\begin{figure}
	\begin{center}
		\includegraphics[width=6in]{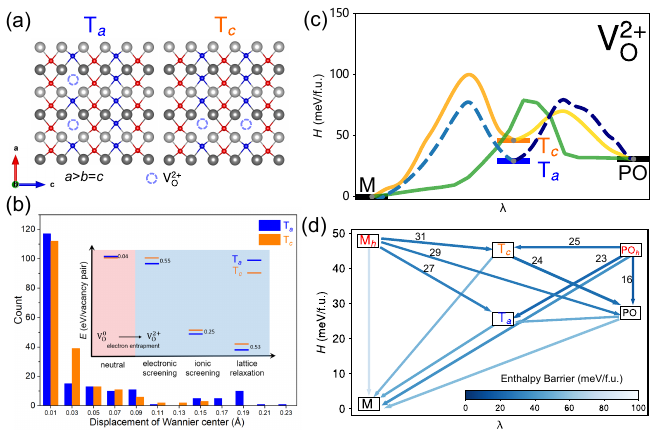}
	\end{center}
	\caption{(a) Schematics of two different configurations of T phase containing a pair of \Vop. 
(b) Histograms of displacements of Wannier centers in T$_a$ and T$_c$ right after the electron entrapment of ${\rm V}_{\rm O}^{0}\rightarrow {\rm V}_{\rm O}^{2+}$. The inset shows different contributions to the change in the relative stability between T$_a$ and T$_c$. Only the energies of configurations containing the same type of oxygen vacancy are comparable.
 (c) Transition pathways in the presence of \Vop~at a concentration of 3.125\%. (d) Phase transition network involving multiple polymorphs of HfO$_2$ containing V$_{\rm O}^{2+}$. The color of the arrow scales with the magnitude of the transition barrier (see values in Table S5). 
	\label{charge}}
\end{figure}

Recently, He~\etal reported that the presence of V$_{\rm O}^{2+}$ beyond some critical concentration will make PO more stable than M~\cite{He21pL180102}. At a low V$_{\rm O}^{2+}$ concentration of 3.125 f.u.\% studied here, we carefully examine the energetics of different configurations containing V$_{\rm O}^{2+}$ and find that M remains to be most stable thermodynamically (See Table S2 in SM\cite{SI}). 
Interestingly, the relative stability between two representative configurations of the T phase, denoted as T$_a$ and T$_c$ (Fig.~\ref{charge}a), depends sensitively on the charge state of oxygen vacancy. Specifically, T$_a$ and T$_c$ has the vacancy pair aligned along the long axis $a$ and the short axis $c$, respectively. Such local oxygen vacancy ordering is almost inevitable statistically even at a low vacancy concentration (see Sec.~\uppercase\expandafter{\romannumeral1}.E and experimental evidences in Sec.~\uppercase\expandafter{\romannumeral3} of SM\cite{SI}). The energies of ${\rm V}_{\rm O}^{0}$@T$_a$ and ${\rm V}_{\rm O}^{0}$@T$_c$ are comparable when the oxygen vacancy is charge neutral, but ${\rm V}_{\rm O}^{2+}$@T$_c$ becomes rather unstable relative to ${\rm V}_{\rm O}^{2+}$@T$_a$ after the electron entrapment of ${\rm V}_{\rm O}^{0}\rightarrow {\rm V}_{\rm O}^{2+}$. The strong impact of \Vop~on the relative stability is confirmed with both 96-atom and 768-atom supercells (Table S3).
{We perform a series of calculations to identify and separate the contributions to the change in the relative stability (see Sec.~\uppercase\expandafter{\romannumeral4} of SM\cite{SI}). As shown in Fig.~\ref{charge}b, the electronic screening already makes ${\rm V}_{\rm O}^{2+}$@T$_a$ containing \Vop~more stable than ${\rm V}_{\rm O}^{2+}$@T$_c$ by 0.55~eV. We compute the displacements of all Wannier centers (WCs, roughly viewed as the positions of electrons, Fig.~S8) after ${\rm V}_{\rm O}^{0}\rightarrow {\rm V}_{\rm O}^{2+}$. The displacements of some WCs in T$_a$ are much larger than those in T$_c$, indicating a more substantial electronic charge density redistribution in T$_a$. The most displaced WC in V$_{\rm O}^{2+}$@T$_a$ is of Hf-$5p$ character (Fig.~S9).  After incorporating the effects of ionic screening and lattice relaxation, {${\rm V}_{\rm O}^{2+}$@T$_a$} containing \Vop~ remains more stable than ${\rm V}_{\rm O}^{2+}$@T$_c$} by 0.53~eV.

The MEPs connecting ${\rm V}_{\rm O}^{2+}$@T$_a$ (${\rm V}_{\rm O}^{2+}$@T$_c$) to the most stable configurations of M and PO phases containing V$_{\rm O}^{2+}$ are shown in Fig.~\ref{charge}c. We find that ${\rm V}_{\rm O}^{2+}$@T$_c$$\rightarrow{\rm V}_{\rm O}^{2+}$@PO with a barrier of 24 meV/f.u. is kinetically favored over ${\rm V}_{\rm O}^{2+}$@T$_c$$\rightarrow{\rm V}_{\rm O}^{2+}$@M with a barrier of 54 meV/f.u., while M and PO remain well separated by a large barrier. Therefore, V$_{\rm O}^{2+}$ promotes T$_c\rightarrow$PO but suppresses T$_c\rightarrow$M. As T$_a$ is more difficult to undergo phase transitions (Fig.~\ref{charge}c), it also explains why as-prepared thin films of HfO$_2$ often exhibit a high volume fraction of nonpolar T phase.  

The V$_{\rm O}^{2+}$-promoted nonpolar-polar structural polymorphism, T$_c\rightarrow$PO, likely plays an important role for the emergence of ferroelectricity. A capping electrode was often required to induce the ferroelectricity in Si-doped HfO$_2$ thin films~\cite{Kim17p242901,Wan22p1805}. This could be due to the easy formation of \Vop~at the metal/HfO$_2$ interface as the two electrons of \Von~may fall to the metal Fermi level~\cite{Cho08p233118}. Since raising the temperature effectively increases the charge-carrier concentration, we propose that during the high-temperature annealing treatment of thin films, the formation of \Vop~at the metal/HfO$_2$ interface is facilitated such that the occurrence of T$_c\rightarrow$PO becomes substantial. Recent observations that the light ion bombardment can greatly enhance the ferroelectricity in HfO$_2$-based thin films~\cite{Kang22p731} corroborates the \Vop-promoted formation of polar HfO$_2$.
The difficulty to obtain the PO phase of HfO$_2$ through bulk synthesis may be attributed to the high formation energy of oxygen vacancy in bulk HfO$_2$ (6.38 eV)~\cite{Robertson07132912}. Moreover, the necessity of \Vop~for T$_c\rightarrow$PO is consistent with a large body of experimental data that nearly all cation dopants inducing ferroelectricity in HfO$_2$-based thin films are of $p$-type because the substitution of Hf$^{4+}$ with acceptor dopants such as Y$^{3+}$ naturally promotes the formation of \Vop~to maintain charge neutrality. The latest verification of intrinsic ferroelectricity in Y-doped HfO$_2$ thin films with a large remnant polarization of 64~$\mu$C/cm$^2$~\cite{Yun22p903} serves as another example supporting our hypothesis. 
For the same reason, the emergence of ferroelectricity in N-doped HfO$_2$~\cite{Xu16p091501} could be arsing from \Vop~that compensates {N$_{\rm O}^-$}~\cite{Umezawa05p143507}.

The finding that \Vop~can strongly modulate the relative configurational stability is not unique to the T phase. We discover an oxygen-deficient configuration of M (denoted as M$_h$) which is stable with \Von~but becomes highly unstable upon the charge entrapment from V$_{\rm O}^{0}$. It is kinetically feasible for ${\rm V}_{\rm O}^{2+}$@M$_h$ to transform to ${\rm V}_{\rm O}^{2+}$@T$_c$/T$_a$ and ${\rm V}_{\rm O}^{2+}$@PO ($<29$~meV/f.u.), highlighting that V$_{\rm O}^{2+}$ can enable a previously forbidden transition of M$\rightarrow$PO. We also identify a configuration of PO, denoted as PO$_h$, which is destabilized by V$_{\rm O}^{2+}$ and easily transforms to nonpolar T and M phases. Contributions to the destabilization effect of \Vop~are separated and quantified for M$_h$ and PO$_h$, revealing a subtle competition between the electronic/ionic screening and lattice relaxation (see Sec.~\uppercase\expandafter{\romannumeral4} of SM\cite{SI}).
The phase transition network involving multiple polymorphs containing V$_{\rm O}^{2+}$ (Fig.~\ref{charge}d) reveals that V$_{\rm O}^{2+}$ can enable facile transitions among nopolar M and T phases and the polar PO phase, the kinetic barriers of which are lower than the polarization switching barrier of the PO phase ($\approx$55~meV/f.u.). We suggest that the charge state of V$_{\rm O}$ is another tuning knob to modulate the stabilities among competing phases in addition to epitaxial strain and electric boundary condition in modern thin-film technology~\cite{Wang22p197601}.

\begin{figure}
	\begin{center}
		\includegraphics[width=6in]{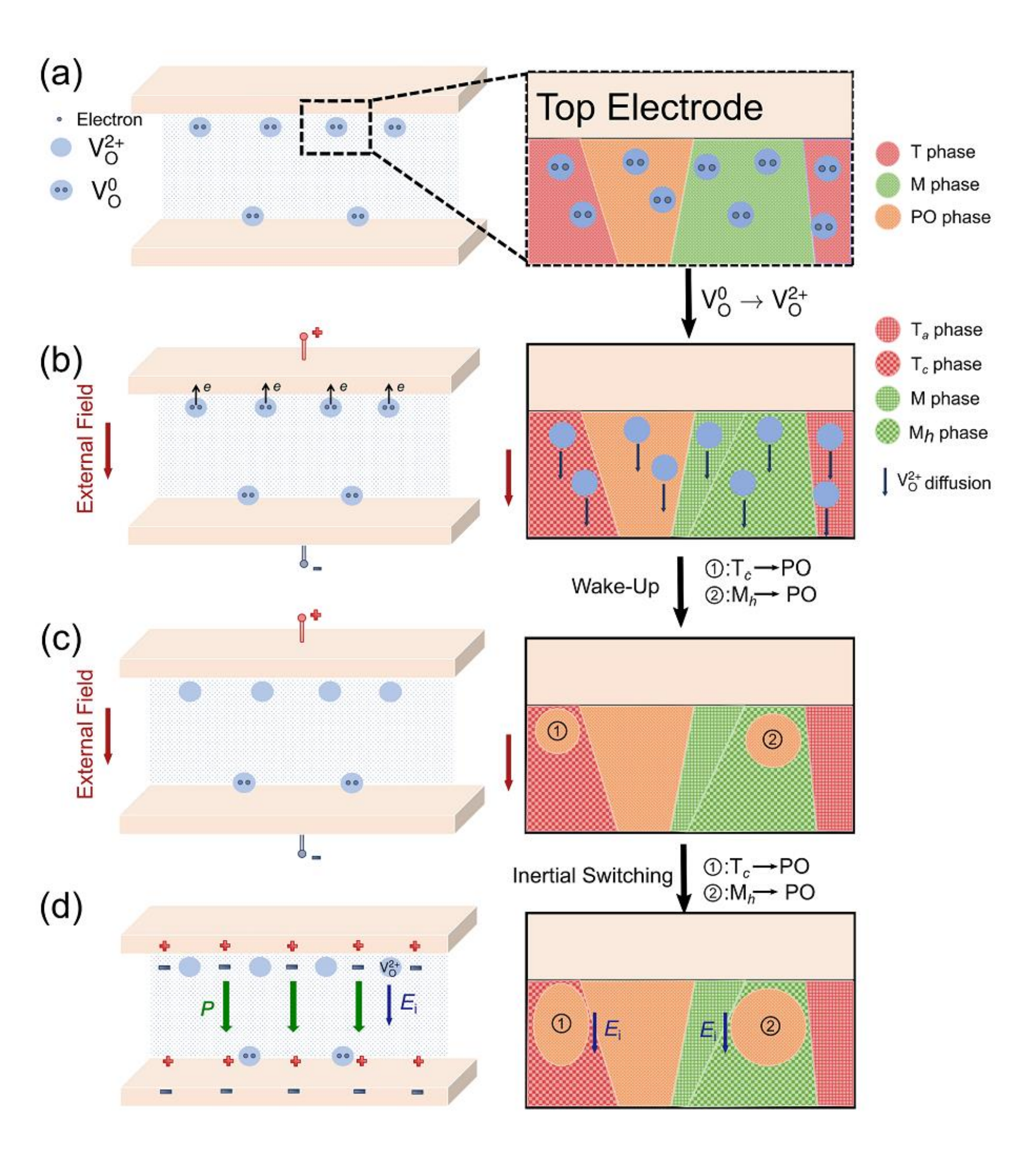}
	\end{center}
	\caption{Schematics of wake-up and inertial switching enabled by \Vop-promoted nonpolar-polar structural polymorphisms. Oxygen vacancies in the right schematics are not shown in (c) and (d) for clarity. 
	\label{exp}}
\end{figure}

The intricate coupling between \Vop~and structural polymorphism kinetics could contribute to performance instability issues including wake-up and imprint. In a pristine HfO$_2$-based capacitor at room temperatures, \Von~is the dominant type of oxygen vacancy, as supported by recent DFT calculations~\cite{Wei21p2104913} and experiments~\cite{Hamouda20p252903}. Therefore, \Von~may distribute at the metal-ferroelectric interface in the matrix of M, T, and PO phases (Fig.~\ref{exp}a). Upon the application of a voltage, electrons of \Von~tend to detrap to the adjacent electrode (top electrode in Fig.~\ref{exp}b), leading to the formation of \Vop~and the activation of M$_h\rightarrow$PO and T$_c\rightarrow$PO (Fig.~\ref{exp}c). The diffusion of \Vop~, recently observed in experiments~\cite{Nukala21p630}, further helps the nonpolar-polar phase transitions in regions away from the electrode. Similar processes will occur near another electrode after reversing the bias.
The overall result is a reduction of M and T and an increase of PO after field cycling, leading to the wake-up effect~\cite{Grimley16p1600173}. 

Finally, M$_h\rightarrow$PO and T$_c\rightarrow$PO to some extent explains  ``fluid imprint" in ferroelectric La-doped HfO$_2$ capacitors where the imprint is easily changeable and has a strong dependence on the switching prehistory~\cite{Buragohain19p35115}.
As illustrated in Fig.~\ref{exp}d, after the application of a preset pulse that polarize an woken-up capacitor (that still has residual nonpolar phases), the electron entrapment occurs more easily  near the negative bound charge side. The presence of \Vop creates a local electric field that aligns with the bulk polarization, a probable source of imprint field ($E_{\rm i}$). Even in the absence of additional bias, the \Vop-promoted nonpolar-polar phase transitions, due to their low kinetic barriers ($<29$~meV/f.u.), could still occur driven by $E_{\rm i}$, leading to continued polarization switching toward the same direction as the previously applied field, a feature of ``inertial switching"~\cite{Buragohain19p35115}. The fluid imprint is associated with an evolving microstructure with a changing volume fraction of the PO phase.

In summary, the robust ferroelectricity of the $Pca2_1$ phase against charge doping highlights the necessity of understanding the effects of charged defects on the ferroelectric properties of HfO$_2$-based devices. 
Our DFT calculations suggest that the positively charged oxygen vacancy can drastically modulate the relative configurational stability of the same phase.
The \Vop-promoted nonpolar-polar structural polymorphism is probably an overlooked yet important mechanism 
for the emergence of polar $Pca2_1$ phase in HfO$_2$-based thin films, offering a new perspective to experimentally observed electrode and doping effects. Generic mechanisms based on the coupling between \Vop~and structural polymorphism kinetics are proposed to explain performance instability issues. To fully understand the wake-up and imprint effects, continued experimental and theoretical efforts are needed. Future studies focusing on the coupling between structural polymorphism kinetics and other extrinsic effects such as strain and direct experimental characterizations of these mechanisms are important for comprehending this complicated ferroelectric compound.     

\begin{acknowledgments}
L.M. and S.L. acknowledge the supports from National Key R\&D Program of China (2021YFA1202100), National Natural Science Foundation of China (12074319), and Westlake Education Foundation. The computational resource is provided by Westlake HPC Center. We thank Dr. Yichun Zhou and Jiangheng Yang for useful discussions inspired by which we discovered the importance of the X$_2^-$ mode for phase transitions.
\end{acknowledgments}

\bibliography{SL.bib}
 
\newpage

\newpage

\newpage

\newpage

\end{document}


\title{Supplemental Material: Structural Polymorphism Kinetics Promoted by Charged Oxygen Vacancies in HfO$_2$}
\author{Li-Yang Ma}
\affiliation{Fudan University, Shanghai 200433, China}
\affiliation{Key Laboratory for Quantum Materials of Zhejiang Province, Department of Physics, School of Science, Westlake University, Hangzhou, Zhejiang 310024, China}
\author{Shi Liu}
\email{liushi@westlake.edu.cn}
\affiliation{Key Laboratory for Quantum Materials of Zhejiang Province, Department of Physics, School of Science, Westlake University, Hangzhou, Zhejiang 310024, China}
\affiliation{Institute of Natural Sciences, Westlake Institute for Advanced Study, Hangzhou, Zhejiang 310024, China}

\date{\today}
\maketitle
\clearpage
\newpage
\section{Hafnia Polymorphs in the presence of oxygen vacancies}
\subsection{Hafnia Polymorphs}

The most stable phase of HfO$_2$ in bulk at ambient conditions is the monoclinic (M) $P2_1/c$ phase. With increasing temperature, the tetragonal (T) $P4_2/nmc$ phase and the cubic $Fm\bar{3}m$ phase will appear sequentially. At a higher pressure ($>$10 GPa), HfO$_2$ transforms to the “antiferroelectric-like” orthorhombic (AO) $Pbca$ phase. All these phases are centrosymmetric. The non-centrosymmetric orthorhombic phase of HfO$_2$ may adopt two polar space groups, $Pca2_1$ and $Pnm2_1$ (denoted as PO and PO$'$). The $Pca2_1$ phase has the polarization aligned along [100] while the $Pnm2_1$ phase has the polarization pointing along [110]. The optimized lattice constants and relative energies are reported in Table~\ref{lattice}. 

\begin{table}[htp]
 	\caption{Lattice parameters in \AA~and relative energies in meV per unit cell (meV/f.u.) in reference to the M phase. \label{lattice}}
 	\begin{center}
 		\begin{tabular}{p{1.5cm}p{2cm}p{2cm}p{2cm}p{2cm}}
 			\hline
 			\hline
 			 & $a$ & $b$ & $c$ & $E$ \\ 
 			\hline
 			T                 & 5.225        & 5.074        & 5.074                  & 166.3                   \\ 
 			$Pca2_1$           & 5.266        & 5.046        & 5.077                  & 84.3                   \\ 
 			$Pmn2_1$           & 5.183        & 5.117        & 5.117                   & 142.9                    \\ 
 			AO                        & 10.469       & 5.073        & 5.074                   & 72.9           \\ 
 			M                         & 5.326        & 5.137          & 5.188                      & 0          \\ 
 			\hline
 			\hline
 		\end{tabular}
 	\end{center}
 \end{table}
 
\clearpage
\newpage
\subsection{Intrinsic phase transition conserving the sign of X$_2^-$}

\begin{figure}[htp]
	\begin{center}
		\includegraphics[width=6in]{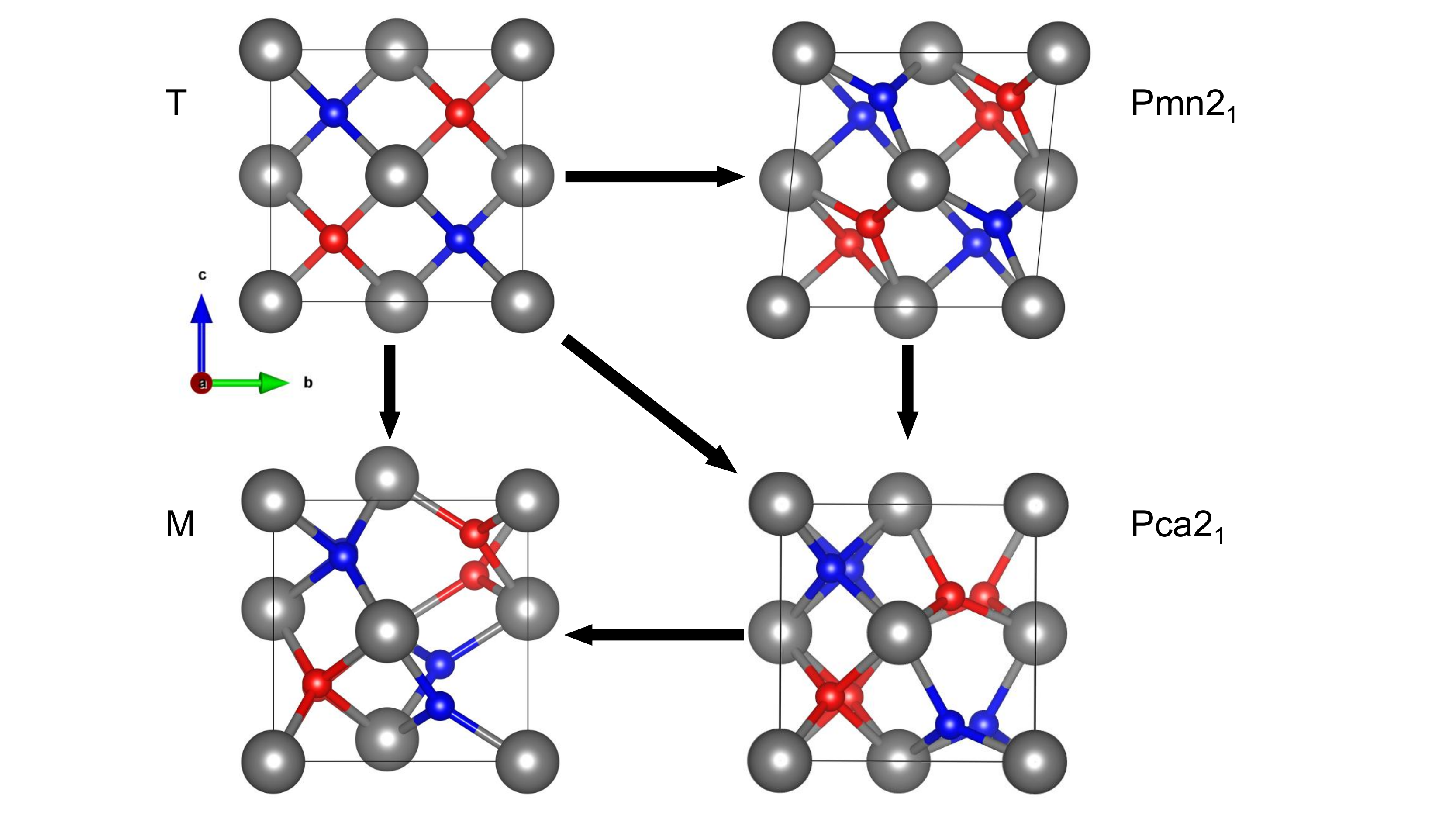}
	\end{center}
	\caption{Transition pathways that conserve the sign of the X$_2^-$ model. The blue and red colors indicate the oxygen atoms that are outward and inward displaced, respectively.\label{intrinsic}}
\end{figure}

In FIG.~\ref{intrinsic}, we illustrate the phase transition pathways between different polymorphs of HfO$_2$. The sign of the X$_2^-$ mode is maintained during the phase transition. The blue and red colors indicate the oxygen atoms that are outward and inward displaced, respectively. Each path is constructed by at least 20 images. When the root-mean-square forces are lower than 0.03 eV/\AA~on images, VCNEB calculations are considered converged. 

\clearpage
\newpage
\subsection{X$_2^-$ and X$_5^y$ modes during the transition of M$\rightarrow$PO}

\begin{figure}[htp]
	\begin{center}
		\includegraphics[width=6in]{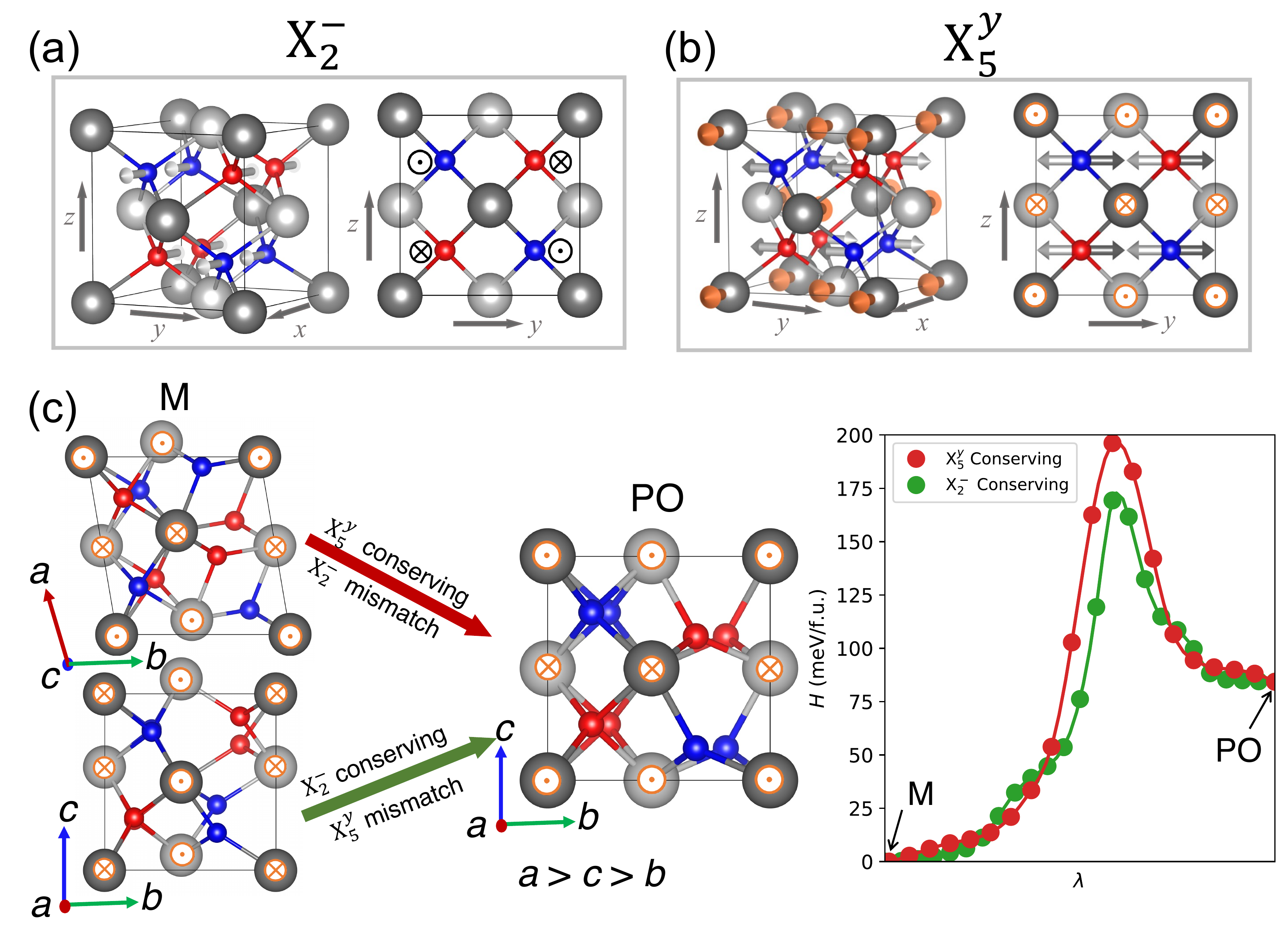}
	\end{center}
	\caption{(a) Schematic illustration of the X$_2^-$ mode in the tetragonal phase of HfO$_2$. The tetragonal unit cell has the long axis along $x$. The closer and further Hf atoms are colored in dark and light grey; outward and inward displaced oxygen atoms are colored in blue and red, respectively. (b) Schematic illustration of the X$_5^y$ mode. Viewing along the $x$-direction, the hafnium atoms along $z$ alternately displace inward and outward. (c) Two possible transition pathways of M$\rightarrow$PO. One pathway conserves the displacement pattern of the X$_5^y$ mode but cannot match the X$_2^-$ mode; another pathway conserves the X$_2^-$ mode but mismatches the X$_5^y$ mode. Both pathways need to overcome large kinetic barriers ($>160$ meV/f.u.). 
 \label{Xmode}}
\end{figure}

As shown in Fig.~\ref{Xmode}a, 
the X$_2^-$ lattice mode is characterized by antiparallel $x$-displacements of neighboring oxygen atoms in the $yz$ plane. We explained in the main text that a transition pathway that conserves the sign of the X$_2^-$ mode has a lower enthalpy barrier than a pathway that reverses the X$_2^-$ mode. Interestingly, the transition of M$\rightarrow$PO involves another lattice mode, X$_5^y$, that is characterized by antiparallel $x$-displacements of Hf atoms and $y$-displacements of oxygen atoms (Fig.~\ref{Xmode}b). 
We find that it is impossible to conserve both X$_2^-$ and X$_5^y$ modes during the transition of M$\rightarrow$PO, as sketched in Fig.~\ref{Xmode}c. The pathway that conserves the displacement pattern of the X$_5^y$ mode fails to match the X$_2^-$ mode; another pathway conserves the X$_2^-$ mode but mismatches the X$_5^y$ mode. Consequently, both pathways possess large kinetic barriers of 169 and 196 meV/f.u., respectively.

\clearpage
\newpage
 \subsection{Modeling oxygen-deficient HfO$_2$}
We remove one oxygen atom from a $2\times2\times1$ supercell containing 32 oxygen atoms to model defective HfO$_{2-x}$ with a vacancy concentration of 3.125 f.u.\%. For a given phase, there exists multiple oxygen-deficient configurations. The configuration is labeled by two indexes, the oxygen coordination number and the vacancy ordering direction. Take the $Pca2_1$ phase ($a>c>b$) as an example. There are six unique configurations for a $2\times2\times1$ supercell containing one oxygen vacancy: PO$_a$-3, PO$_b$-3, and PO$_c$-3 that has a threefold-coordinated oxygen removed,  and PO$_a$-4, PO$_b$-4, and PO$_c$-4 that has a fourfold-coordinated oxygen removed. The lattice constants and atomic positions are fully optimized.
The relative energies of all configurations are reported in Table~\ref{defect}. All files of optimized structures are uploaded to \url{https://github.com/sliutheorygroup/supplementary-material/tree/main/L23}.

\begin{figure}[htp]
	\begin{center}
		\includegraphics[width=6in]{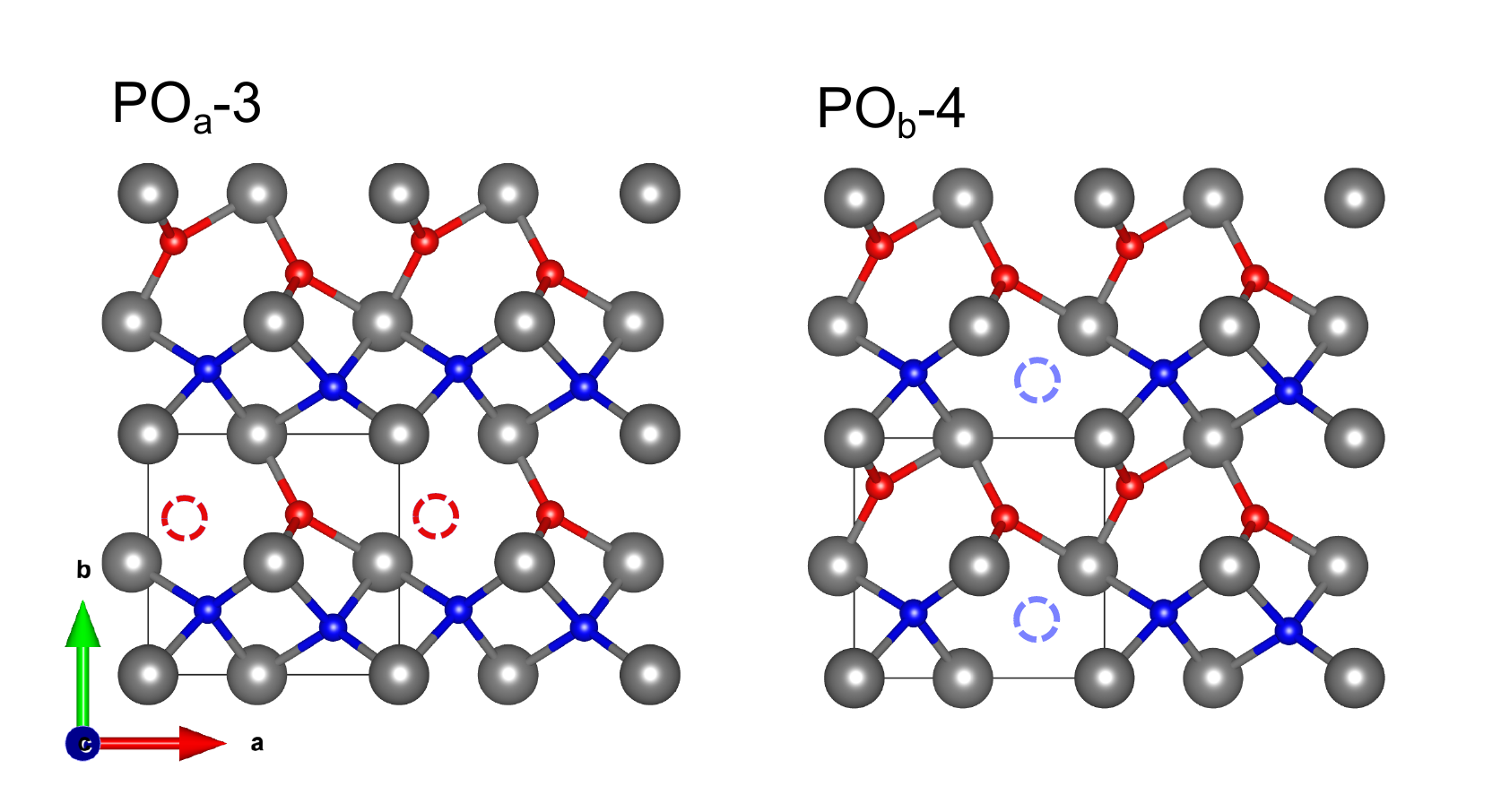}
	\end{center}
	\caption{Schematics of PO$_a$-3 and PO$_b$-4.\label{POphase}}
\end{figure}

\begin{table}[htp]
\centering
\caption{Energy of oxygen-deficient HfO$_2$ containing V$_{\rm O}^0$ and V$_{\rm O}^{2+}$ in meV per formula unit (meV/f.u.) relative to the lowest-energy configuration. M$_a$-4 (M$_b$-4) is the M$_h$ configuration; PO$_a$-4 (PO$_c$-4) is the PO$_h$ configuration discussed in the main text and Fig.~3d.\label{defect}}

\begin{tabular}{p{2cm}p{2cm}|p{2cm}|p{2cm}}
\hline
\multicolumn{2}{c|}{ \multirow{2}*{Configuration} }& \multicolumn{2}{c}{$E$}\\
\cline{3-4}
\multicolumn{2}{c|}{}&V$_{\rm O}^0$&V$_{\rm O}^{2+}$\\
\hline
\multirow{2}*{T}&T$_a$&152.4&32.1\\
\cline{2-4}
&T$_c$&151.1&48.8\\
\hline
\multirow{6}*{$Pca2_1$}&PO$_a$-3&90.6&42\\
\cline{2-4}
&PO$_b$-3&89.8&48.3\\
\cline{2-4}
&PO$_c$-3&88.8&35.2\\
\cline{2-4}
&PO$_a$-4&72.6&51.4\\
\cline{2-4}
&PO$_b$-4&72.7&43.8\\
\cline{2-4}
&PO$_c$-4&75.1&57.5\\
\hline
\multirow{4}*{$Pmn2_1$}&PO$'_a$-3&137.1&31.6\\
\cline{2-4}
&PO$'_c$-3&138.7&44.4\\
\cline{2-4}
&PO$'_a$-4&136.4&31.4\\
\cline{2-4}
&PO$'_c$-4&136.3&43.3\\
\hline
\multirow{6}*{M}&M$_a$-3&8.6&0\\
\cline{2-4}
&M$_b$-3&7.7&2.2\\
\cline{2-4}
&M$_c$-3&8.4&4.3\\
\cline{2-4}
&M$_a$-4&0.9&50.8\\
\cline{2-4}
&M$_b$-4&0&48.6\\
\cline{2-4}
&M$_c$-4&0&4.1\\
\hline
\end{tabular}
\end{table}
\clearpage
\newpage

\subsection{Oxygen vacancy pair in the T phase}

\begin{figure}[htp]
	\begin{center}
		\includegraphics[width=6in]{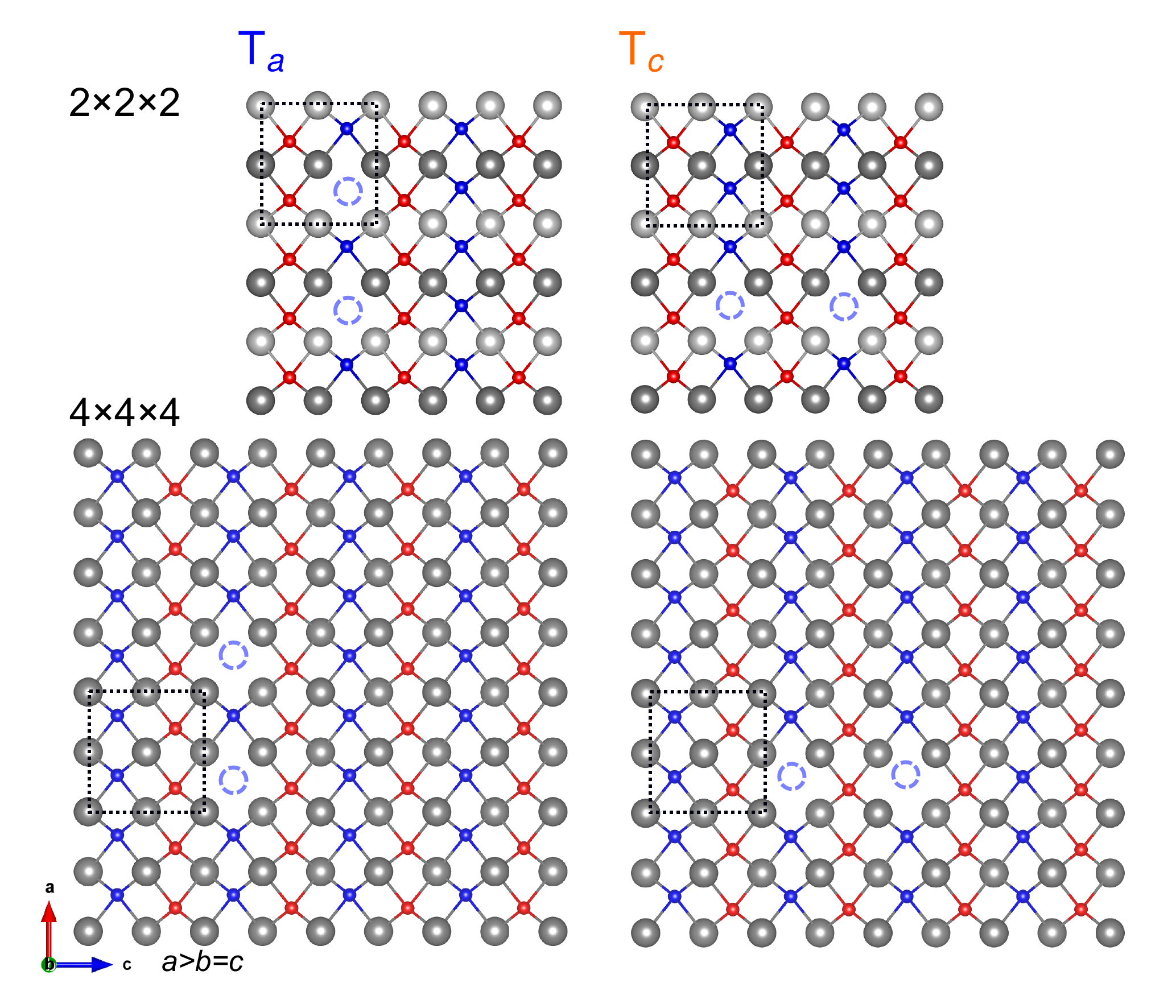}
	\end{center}
	\caption{Schematics of T$_a$ and T$_c$ modeled with a 2$\times$2$\times$2 supercell (top) and a 4$\times$4$\times$4 supercell (bottom).\label{Tphase}}
\end{figure}

We investigated the impact of oxygen vacancy pair on the relative stability between T$_a$ and T$_c$ using supercells of different sizes. The lattice constants are fixed to the ground-state values of the defect-free T phase. The atomic positions are fully relaxed. The structures are shown in FIG.~\ref{Tphase}. As reported in Table~\ref{defectenergy}, once the charge state of the oxygen vacancy becomes +2, the T$_c$ phase becomes highly unstable. 

\begin{figure}[htp]
	\begin{center}
		\includegraphics[width=6in]{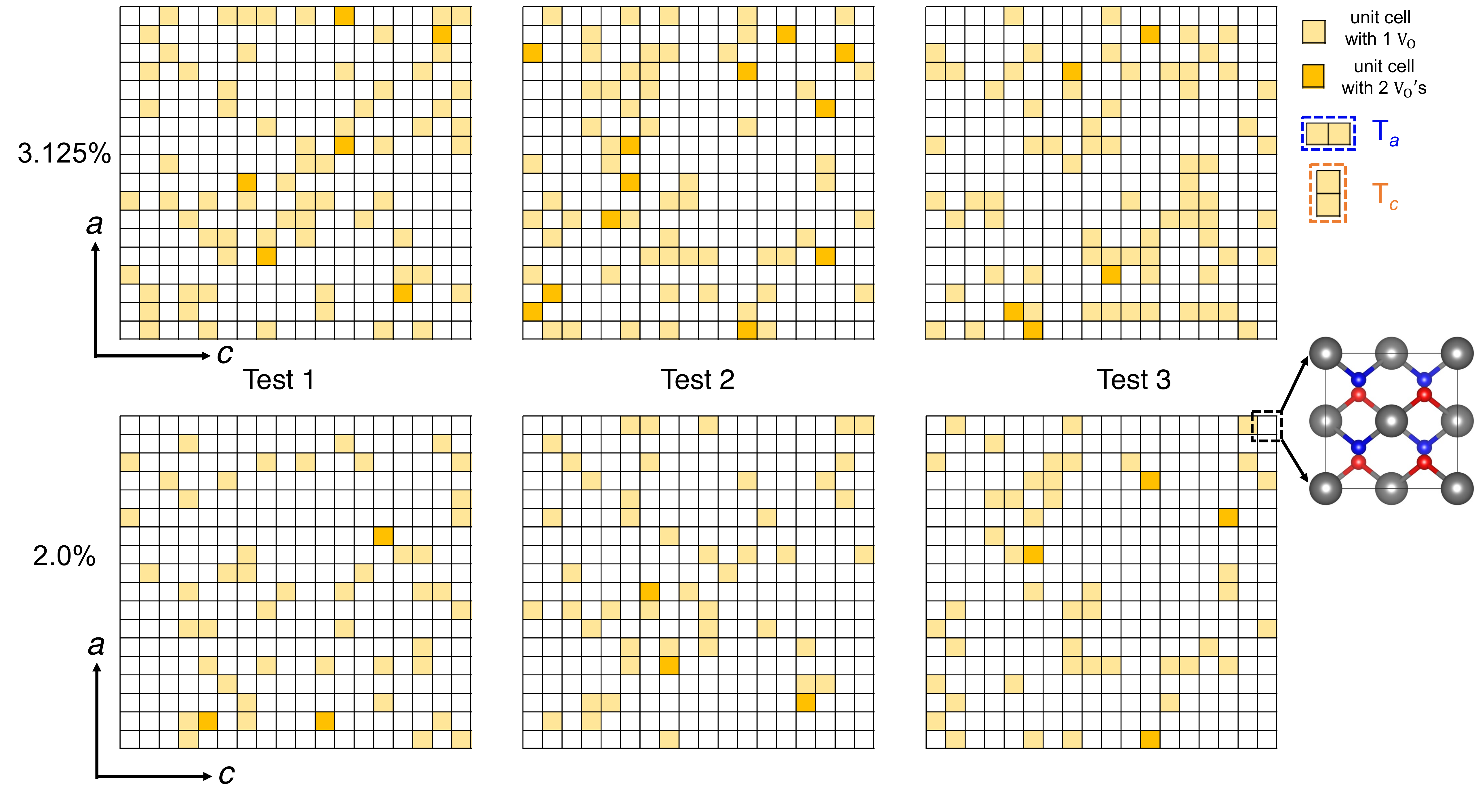}
	\end{center}
	\caption{Oxygen-vacancy distributions in a $18\times18\times18$ supercell at a vacancy concentration of 3.125\% and 2\% from three statistical tests. Only one layer of the supercell is shown for a test. The presence of T$_a$/T$_c$ is inevitable even at a relatively low vacancy concentration.\label{Model}}
\end{figure}

We note that the key structural feature of T$_a$ (T$_c$) is the presence of oxygen vacancies in neighboring unit cells. Statistically, such oxygen vacancy pair is almost {\bf \em inevitable} even at a relatively low vacancy concentration (\ie~3.125\% studied in this work). We performed a simple statistical test where a 18$\times$18$\times$18 T-phase supercell has oxygen vacancies randomly distributed. The distributions of oxygen vacancies from multiple tests are illustrated in Fig.~\ref{Model}. It is evident that many oxygen vacancy pairs adopt T$_a$/T$_c$ ordering locally even at a low concentration of 2.0\%.

\begin{table}[htbp]
\centering
\caption{Energies of supercells containing oxygen vacancy pair in the T phase. Energy in eV. \label{defectenergy}}
\begin{tabular}{p{3cm}|p{3cm}|p{2cm}p{2cm}p{2cm}}
\hline
 
\multirow{2}*{Supercell}& \multirow{2}*{Charge State} &\multicolumn{3}{c}{$E$}\\
\cline{3-5}
&&T$_a$&T$_c$&T$_c$-T$_a$\\
\cline{1-5}
\multirow{2}*{2$\times$2$\times$2}&V$_{\rm O}^0$&-949.282&-949.334&-0.0519\\
\cline{2-5}
&V$_{\rm O}^{2+}$&-975.887&-975.369&0.518\\
\hline
\multirow{2}*{4$\times$4$\times$4}&V$_{\rm O}^0$&-7749.796&-7749.809&-0.0135\\
\cline{2-5}
&V$_{\rm O}^{2+}$&-7777.126&--7776.621&0.505\\
\hline
\end{tabular}
\end{table}

\clearpage
\newpage
\section{Structural polymorphisms in the presence of oxygen vacancies}

We explored all possible transition pathways that conserve the sign of the X$_2^-$ mode connecting low-energy oxygen-deficient configurations in the presence of V$_{\rm O}^0$ and V$_{\rm O}^{2+}$, respectively. The results are reported in Table~\ref{pathsVo} and Table~\ref{pathsVop}. All converged transition pathways can be found in \url{https://github.com/sliutheorygroup/supplementary-material/tree/main/L23}.

\begin{table}[htp]
\centering
\caption{Phase transition barriers in the presence of V$_{\rm O}^0$ in meV per formula unit (meV/f.u.). \label{pathsVo}}

\begin{tabular}{p{1.5cm}p{2.5cm}|p{2cm}|p{2cm}}
\hline
\multicolumn{2}{c|}{ \multirow{2}*{Phase transition} }& \multicolumn{2}{c}{V$_{\rm O}^0$}\\
\cline{3-4}
\multicolumn{2}{c|}{}&{forward}&{reverse}\\
\hline
\multirow{2}*{T$\rightarrow$PO}&T$_a$$\rightarrow$PO$_a$-4&4.4&84.2\\
\cline{2-4}
&T$_c$$\rightarrow$PO$_c$-4&5.9&82\\
\hline
\multirow{2}*{T$\rightarrow$M}&T$_a$$\rightarrow$M$_a$-4&2.7&154.4\\
\cline{2-4}
&T$_c$$\rightarrow$M$_c$-4&5.9&157.1\\
\hline
\multirow{1}*{M$\rightarrow$PO}&M$_b$-4$\rightarrow$PO$_b$-4&183.9&111\\
\hline
\end{tabular}
\end{table}

\begin{table}[htp]
\centering
\caption{Phase transition barriers in the presence of V$_{\rm O}^{2+}$ in meV per formula unit (meV/f.u.). M$_a$-4 (M$_b$-4) is the M$_h$ configuration; PO$_a$-4 (PO$_c$-4) is the PO$_h$ configuration discussed in the main text. The values colored in blue are used to plot Fig.~3d. \label{pathsVop}}

\begin{tabular}{p{2cm}p{3cm}|p{2cm}|p{2cm}}
\hline
\multicolumn{2}{c|}{ \multirow{2}*{Phase transition} }& \multicolumn{2}{c}{V$_{\rm O}^{2+}$}\\
\cline{3-4}
\multicolumn{2}{c|}{}&{forward}&{reverse}\\
\hline
\multirow{4}*{fate of T}& \textcolor{blue}{T$_a$$\rightarrow$M$_a$-3}& \textcolor{blue}{48.3}&81.6\\
\cline{2-4}
&\textcolor{blue}{T$_a$$\rightarrow$PO$_c$-3}&50.2&\textcolor{blue}{48.6}\\
\cline{2-4}
&{ \textcolor{blue}{T$_c$$\rightarrow$PO$_c$-3}}&{ \textcolor{blue} {24}}&39.2\\
\cline{2-4}
&T$_c$$\rightarrow$PO$_b$-4&18.4&25.4\\
\cline{2-4}
& \textcolor{blue}{T$_c$$\rightarrow$M$_c$-4}& \textcolor{blue}{53.9}&100\\
\hline

\multirow{1}*{fate of PO}& \textcolor{blue}{PO$_c$-3$\rightarrow$M$_c$-3}& \textcolor{blue}{56.7}&87.5\\
\hline
\multirow{5}*{fate of PO$_h$}& \textcolor{blue}{PO$_a$-4$\rightarrow$T$_a$}& \textcolor{blue}{23.1}&40.7\\
\cline{2-4}
& \textcolor{blue}{PO$_a$-4$\rightarrow$T$_c$}& \textcolor{blue}{25.1}&25.8\\
\cline{2-4}
&\textcolor{blue}{PO$_a$-4$\rightarrow$M$_b$-3}&\textcolor{blue}{35.9}&84.9\\
\cline{2-4}
& \textcolor{blue}{PO$_a$-4$\rightarrow$PO$_c$-3}& \textcolor{blue}{74.1}&90.1\\
\cline{2-4}
&\textcolor{blue}{PO$_c$-4$\rightarrow$PO$_c$-3 (via V$_{\rm O}^{2+}$diffusion)}& \textcolor{blue}{16.1}&38.1\\
\hline
\multirow{5}*{fate of M$_h$}& \textcolor{blue}{M$_a$-4$\rightarrow$T$_a$}& \textcolor{blue}{31.1}&47.9\\
\cline{2-4}
& \textcolor{blue}{M$_a$-4$\rightarrow$T$_c$}& \textcolor{blue}{26.7}&26.8\\
\cline{2-4}
&M$_b$-4$\rightarrow$PO$_b$-4&92.1&97.4\\
\cline{2-4}
& \textcolor{blue}{M$_b$-4$\rightarrow$PO$_c$-3}& \textcolor{blue}{29.3}&42.8 \\
\cline{2-4}
& \textcolor{blue}{M$_a$-4$\rightarrow$M$_a$-3 (via V$_{\rm O}^{2+}$diffusion)}& \textcolor{blue}{81.3}&131.4\\
\hline
\end{tabular}
\end{table}

\clearpage
\newpage
\section{Experiment-Theory Comparison}

Even before the discovery of ferroelectricity in HfO$_2$-based thin films, nonpolar HfO$_{2-x}$ has already been actively studied as a promising resistive switching (RS) material for nonvolatile resistive random access memory (ReRAM)~\cite{Fang13p92,Quinteros14p024501,Duncan16p400,Hou16p123106,Singh18p014501,Antad21p54133}.

\begin{figure}[htp]
	\begin{center}
		\includegraphics[width=6in]{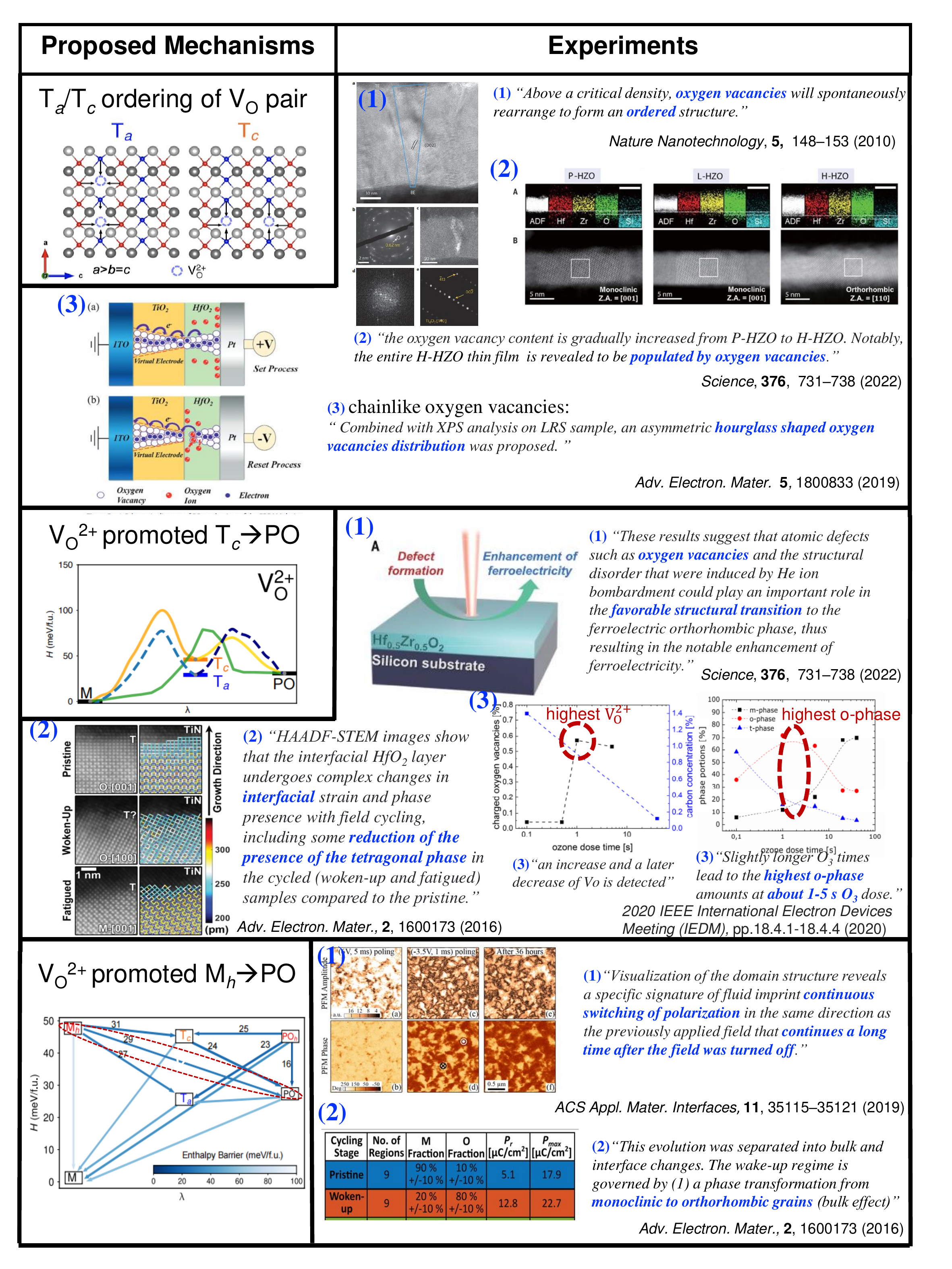}
	\end{center}
	\caption{Experimental evidences supporting various \Vop-dependent mechanisms proposed in this work\cite{Kwon10p148,Kang22p731,Zhang19p1800833,Grimley16p1600173,Mittmann20p18.4.1,Buragohain19p35115}.\label{Exp}}
\end{figure}

The RS mechanism in binary-oxide-based ReRAM has been the subject of numerous of studies~\cite{Clima12p133102,Kamiya13p155201,Duncan16p400,Dirkmann18p14857}. The current consensus is that the ON-OFF resistive switching is governed by the reversible formation and disruption of conducting filaments of {\bf \em chainlike oxygen vacancies}, as illustrated in Fig.~\ref{Exp}.  In particular, Kwon~\etal used high-resolution transmission electron microscopy to probe directly the nanofilaments in a Pt/TiO$_2$/Pt system during RS and confirmed that the switching occurs by the formation and disruption of Ti$_n$O$_{2n-1}$, a phase with {\bf \em ordered} oxygen vacancies~\paper{\cite{Kwon10p148}}. These previous studies of the RS mechanism in binary oxides could be viewed as ``direct" evidences supporting our proposed T$_a$ and T$_c$ ordering in HfO$_2$. 

Third, as summarized in Fig.~\ref{Exp}, our proposed generic mechanisms based on the coupling between \Vop~and structural polymorphism kinetics can well explain a broad range of experimental observations relevant to the origin of ferroelectricity in HfO$_2$-based thin films as well as {\bf \em interfacial phase transitions} at the metal/HfO$_2$ interface, wake-up, fluid imprint, and inertial switching.
For example, the proposed V$_{\rm O}^{2+}$-promoted nonpolar-polar structural polymorphism explains a very recent observation that the light ion bombardment greatly enhanced the ferroelectricity in HfO$_2$-based ferroelectric thin films~\cite{Kang22p731}.
These experimental observations in HfO$_2$-based thin films could also be viewed as ``indirect evidences".

\clearpage
\newpage
\section{Destabilization effects arising from the electron entrapment}

A series of calculations are carried out to fully disentangle the destabilization effects arsing from the electron entrapment of ${\rm V}_{\rm O}^{0}\rightarrow {\rm V}_{\rm O}^{2+}$. We perform three successive calculations in order to identify and separate the contributions to the change in the relative configurational stability. The first calculation is simply a self-consistent field calculation (\ie, single-point energy calculation) without structural relaxations upon the electron entrapment of ${\rm V}_{\rm O}^{0}\rightarrow {\rm V}_{\rm O}^{2+}$. The change in the relative stability can thus be attributed to the difference in {\bf \em electronic screening}. In the second calculation, we relax internal atomic positions with lattice constants fixed. The change in the relative stability is due to varying degrees of {\bf \em ionic screening}. Finally, we perform a full structural optimization by relaxing both internal coordinates and lattice parameters, and name this contribution to the stability change as ``{\bf \em lattice relaxation}". 

The naming scheme used to distinguish different oxygen-deficient configurations is detailed in Sec.~\uppercase\expandafter{\romannumeral1}.D. The configuration is labeled by two indexes, the oxygen coordination number and the vacancy ordering direction. For example, M$_a$-3 denotes a configuration of M phase that has threefold-coordinated oxygen atoms removed, resulting in an oxygen vacancy pair aligned along the $a$ axis. 

\begin{figure}[htp]
	\begin{center}
		\includegraphics[width=6in]{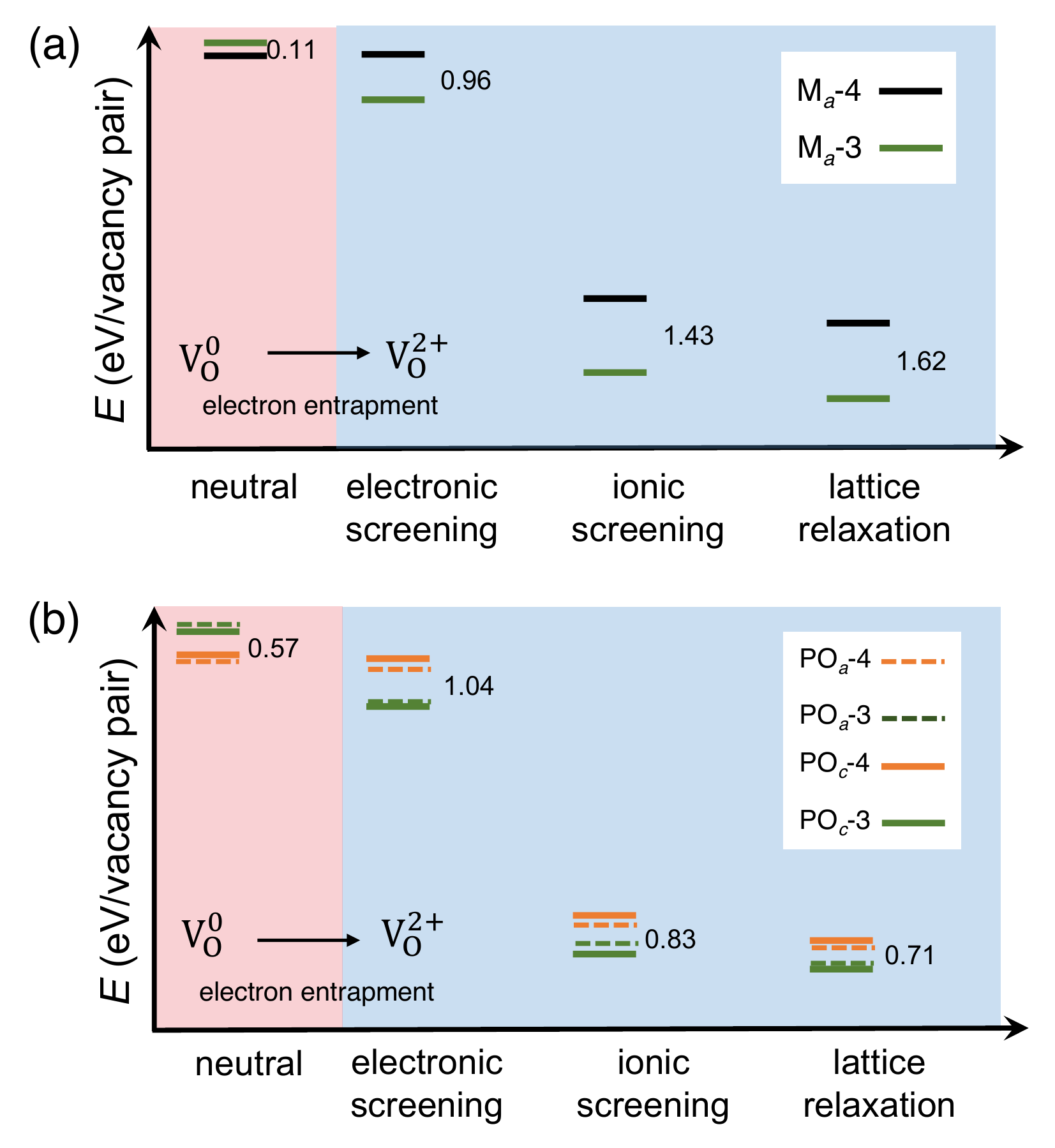}
	\end{center}
	\caption{Different contributions to the change in the relative stability upon the electron entrapment of ${\rm V}_{\rm O}^{0}\rightarrow {\rm V}_{\rm O}^{2+}$ in (a) M phase and (b) PO phase.\label{M_PO}}
\end{figure}

As shown in Fig.~\ref{M_PO}a, in the presence of \Von, M$_a$-4 and M$_a$-3 are close in energy with M$_a$-4 being slightly more stable. We note that M$_a$-4, M$_b$-4, and M$_c$-4 are close in energy, while M$_a$-3, M$_b$-3, and M$_c$-3 have comparable energies (see values in Table S2 in SM). After the electron entrapment from \Von, the electronic screening in M$_a$-3 is stronger such that M$_a$-3 becomes more stable than M$_a$-4 (denoted as M$_h$ with \Vop~in Fig.~3d of the main text) by $\approx$1.0 eV. The ionic relaxation further stabilize M$_a$-3 more than M$_a$-4. Eventually, M$_a$-4 (M$_h$) is higher in energy than the most stable configuration of M$_a$-3 by 1.62 eV in the presence of \Vop. 

The situation is more complicated for the polar orthorhombic (PO) phase (Fig.~\ref{M_PO}b). The lowest-energy configuration with \Von~is PO$_a$-4 while PO$_a$-3 is higher in energy by 0.57 eV; PO$_a$-4 and PO$_c$-4 are close in energy, and PO$_a$-3 and PO$_c$-3 have comparable stability. Right after the electron entrapment of ${\rm V}_{\rm O}^{0}\rightarrow {\rm V}_{\rm O}^{2+}$, the electronic screening in PO$_c$-3 is the strongest, whereas PO$_c$-4 (denoted as PO$_h$ with \Vop~in Fig.~3d of the main text) becomes unstable and has an energy higher by $\approx$1.0 eV, comparable with the energy difference between M$_a$-4 and M$_a$-3 containing \Vop when only the electronic screening is accounted for. We find that both the ionic screening and lattice relaxation will reduce the energy difference between PO$_c$-4 and PO$_c$-3 to $\approx$0.7~eV. Therefore, because of the different degrees of ionic screening and lattice relaxation, the destabilization effect of the charged vacancy in the M phase is much more substantial than that in the PO phase.

We projected the band structures onto the Wannier function (WF) corresponding to the most displaced WC after the electron entrapment of ${\rm V}_{\rm O}^{0}\rightarrow {\rm V}_{\rm O}^{2+}$ in T$_a$ (Fig.~\ref{WF}, left panel).
It turns out the most displaced WC is of Hf-5$p$ character based on the computed projected density of states (PDOS, Fig.~\ref{WF}, right panel). The inset also compares the change in the position of the most displaced WC (represented as a green ball) before and after the electron entrapment. These results reveal an interesting effect of Hf semicore $5p$ electrons on the electronic screening. We note that despite the nearly identical ionic radii of Zr$^{4+}$ and Hf$^{4+}$, many experiments highlighted the drastic differences between ZrO$_2$ and HfO$_2$ concerning the easiness of stabilizing the PO phase in HfO$_2$ but not in ZrO$_2$~\cite{Fan16p012906}. Our finding actually hints at the importance of semicore $5p$ electrons of Hf. With that said, to fully resolve this puzzle demands further investigations beyond the scope of current work.

\begin{figure}[htp]
	\begin{center}
		\includegraphics[width=6in]{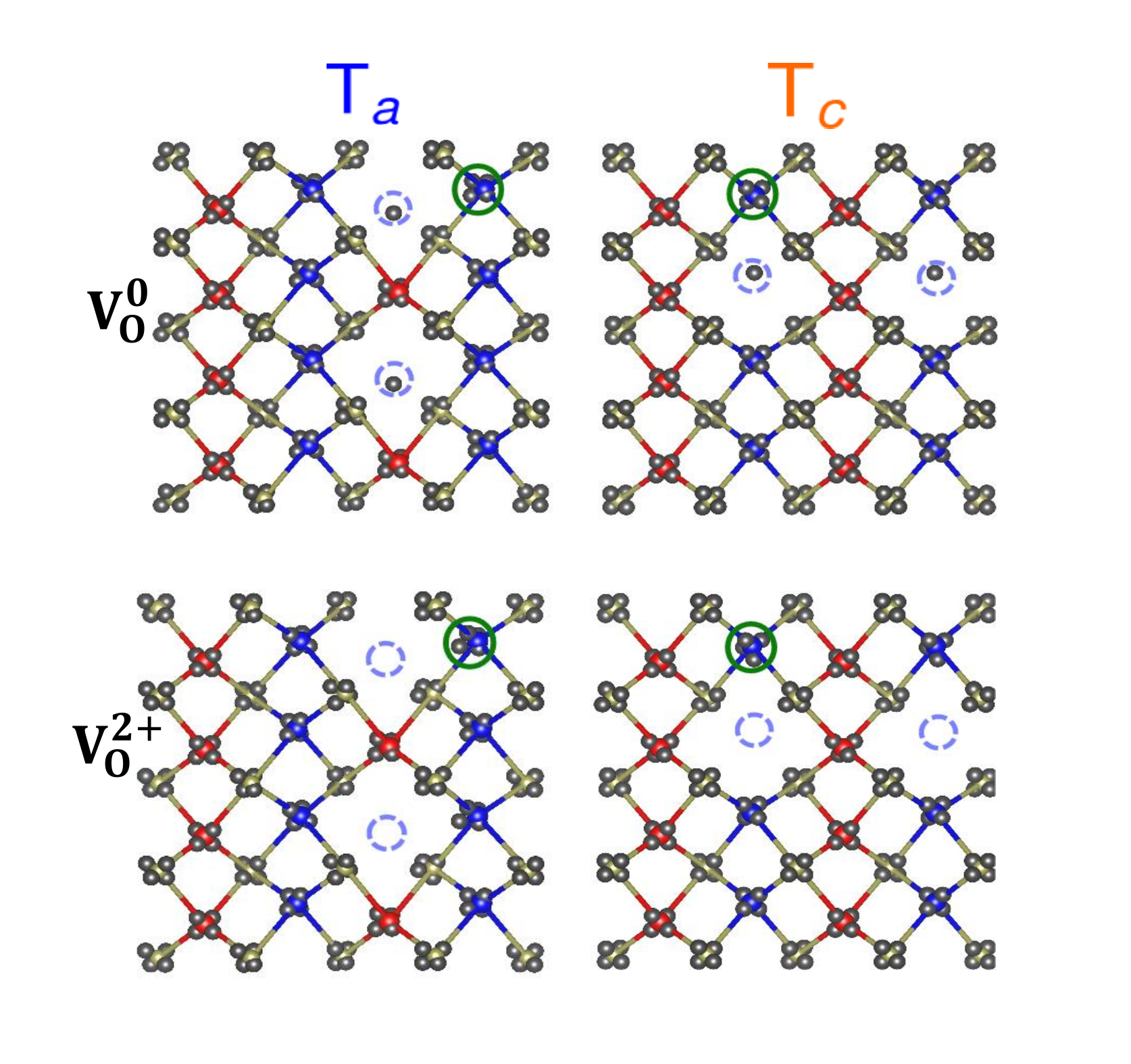}
	\end{center}
	\caption{Wannier centers represented as gray balls in T$_a$ and T$_c$ containing \Von~and \Vop, respectively.\label{WCs}}
\end{figure}

\begin{figure}[htp]
\centering
\includegraphics[width=5in]{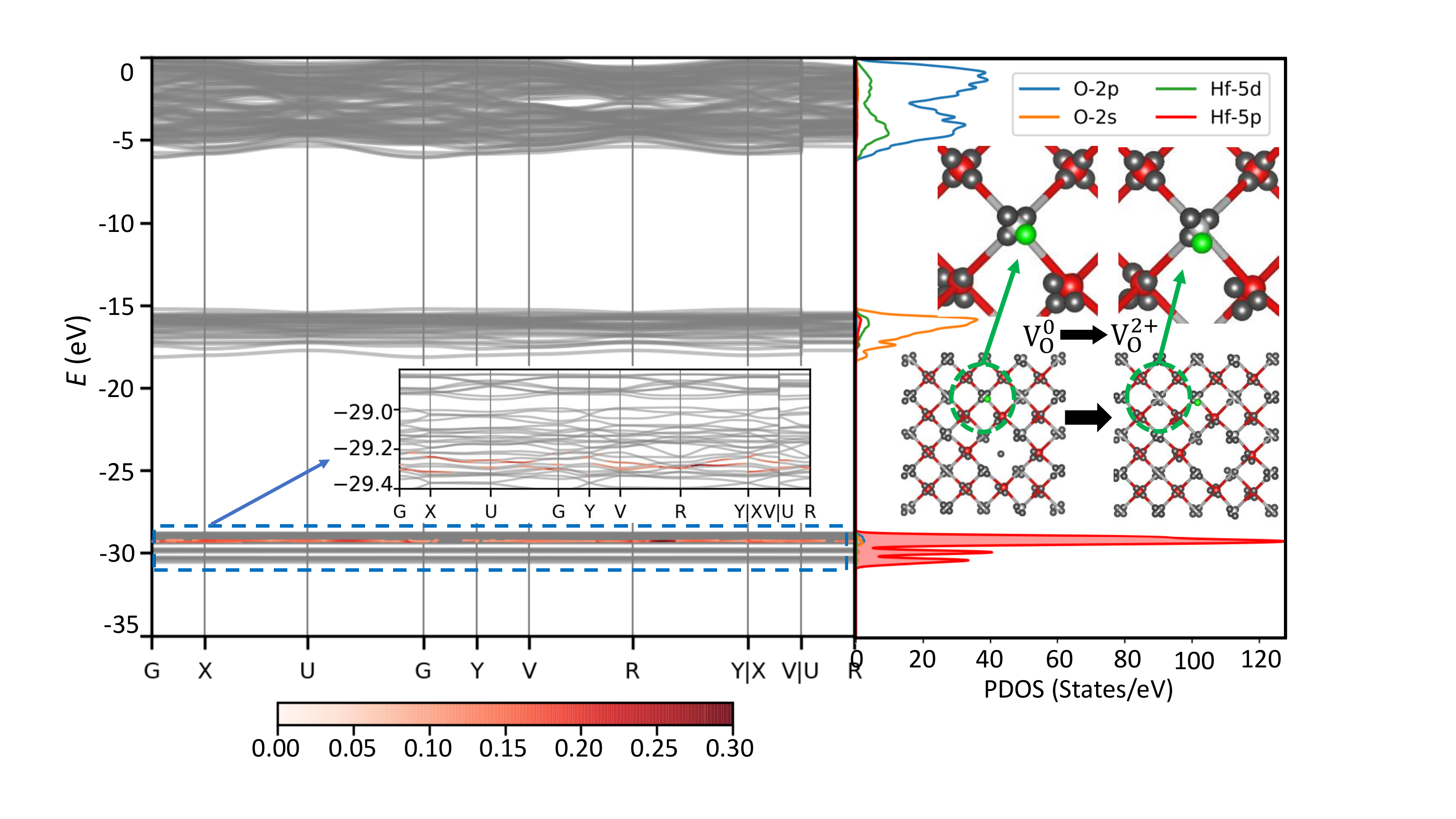}
 \caption{Electronic band structure of ${\rm V}_{\rm O}^{2+}$@T$_a$ with contributions from the most displaced Wannier center colored according to the colorbar. To improve the clarity, only weights larger than 0.1 are shown. The right panel shows the projected density of states (PDOS). 
}
 \label{WF}
 \end{figure}

\clearpage
\newpage
\section{Oxygen vacancy migration}

\begin{figure}[htp]
	\begin{center}
		\includegraphics[width=6in]{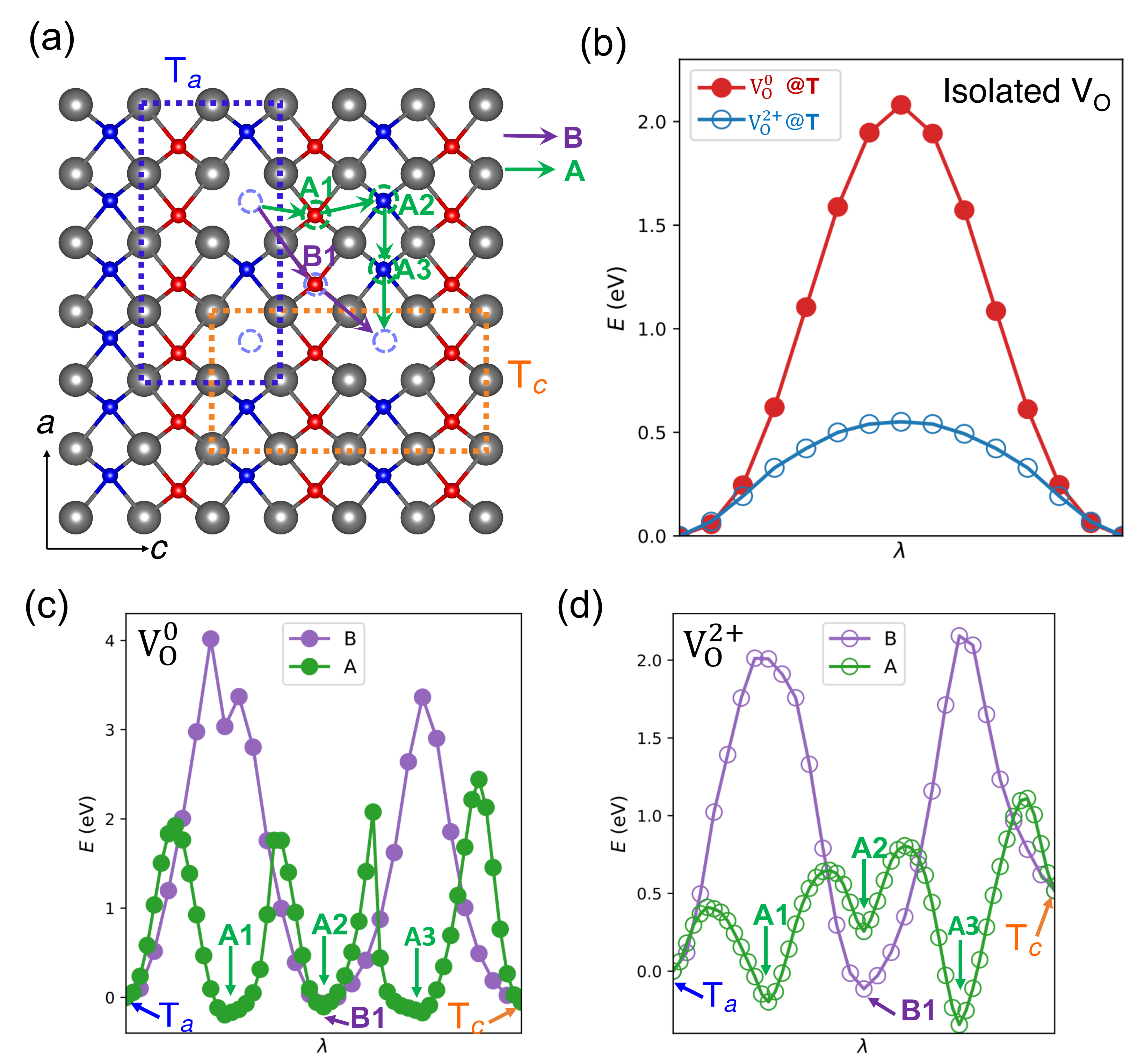}
	\end{center}
	\caption{(a) Possible oxygen vacancy migration pathways (A and B) that transform T$_a$ to T$_c$. Minimum energy pathways obtained with NEB for (b) single V$_{\rm O}^0$/V$_{\rm O}^{2+}$ hopping to nearest-neighbor sites, (c) V$_{\rm O}^0$ migration from T$_a$ to T$_c$, and (d) V$_{\rm O}^{2+}$ migration from T$_a$ to T$_c$.\label{Vo}}
\end{figure}

As shown in Fig.~\ref{Vo}b, the migration barrier of \Von~hopping to the nearest-neighbor site ($\approx2.1$~eV) is much higher than that of~\Vop ($\approx0.4$~eV), consistent with previous DFT investigations~\cite{He21pL180102}. This also indicates the reversible oxygen migration in hafnia-based ferroelectric 
thin films recently observed by the scanning transmission electron microscopy~\paper{\cite{Nukala21p630}} is due to ~\Vop, corroborating our proposed \Vop-related mechanisms.  

We investigated two possible oxygen-vacancy migration pathways that transform T$_a$ to T$_c$ (denoted as A and B in Fig.~\ref{Vo}a), and the corresponding minimum energy pathways (MEPs) determined with DFT-based NEB are reported in Fig.~\ref{Vo}c-d. It is clear that \Vop~is more mobile than \Von. A full transformation from T$_c$ to T$_a$ requires successive oxygen-vacancy hopping, a multistep process that has to overcome multiple barriers. For this reason, we suggest the transition between T$_a$ and T$_c$ via the \Von/\Vop~diffusion is not likely a major process at room temperatures. 

\clearpage
\newpage

\section{Influence of metal electrodes}
\begin{figure}[htp]
\centering
\includegraphics[width=5in]{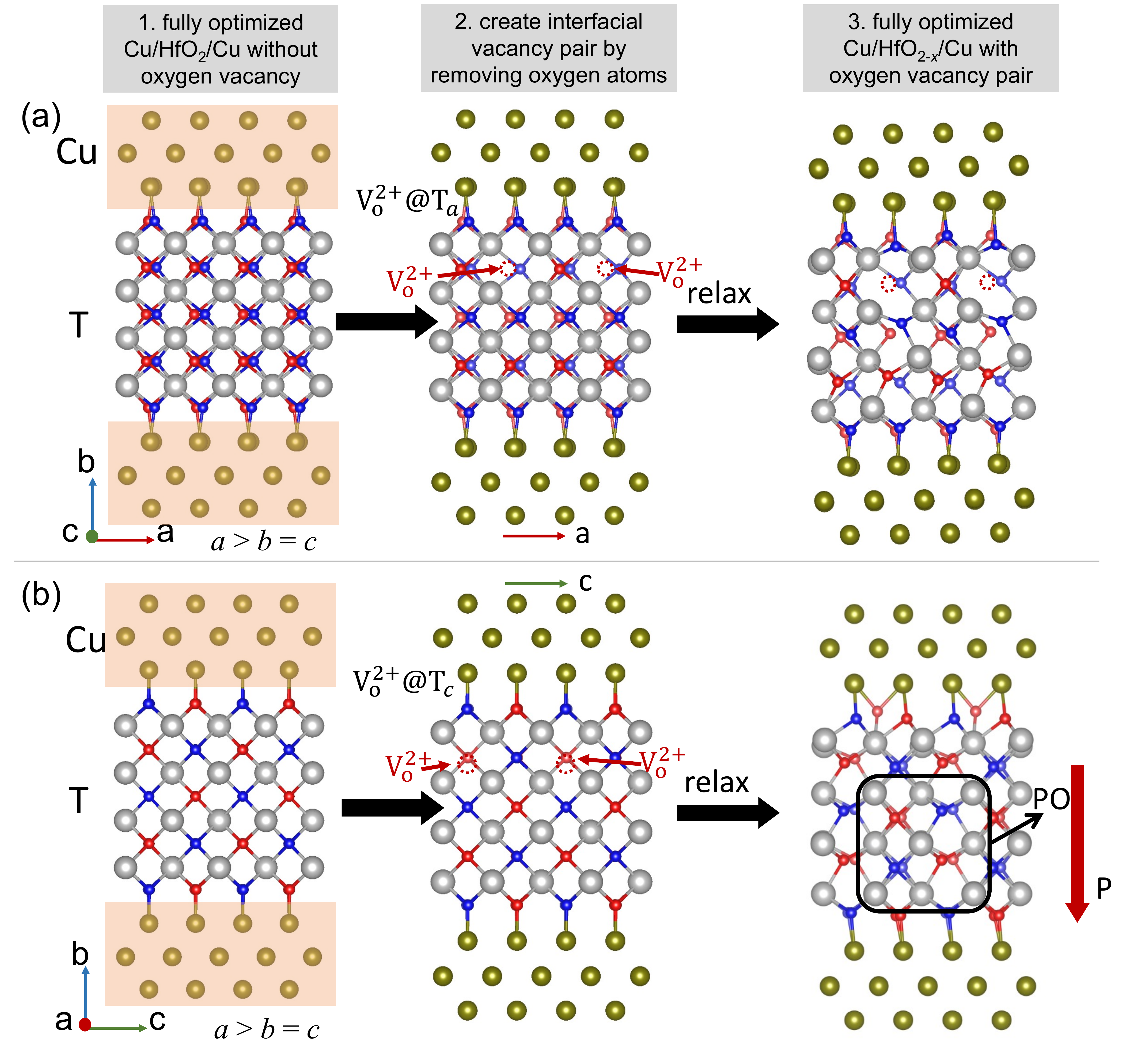}
 \caption{Effects of interfacial oxygen vacancies on the structural transition in a T-phase thin film sandwiched by Cu electrodes. (a) \Vop~pair aligned along $a$ representing \Vop@T$_a$. (b) \Vop~pair aligned along $c$ representing \Vop@T$_c$. Some unit cells transform to the PO phase spontaneously after the structural relaxation.
}
 \label{interface}
 \end{figure}

We performed additional model calculations to understand the effect of interfacial oxygen vacancies on the structural transition in a T-phase thin film sandwiched by Cu electrodes. The Cu electrode is chosen for its small lattice mismatch ($<1.73\%$) with the T-phase of HfO$_2$. As shown in Fig.~\ref{interface}, we first optimized the capacitor-like structure without oxygen vacancy. Then we introduced the V$_{\rm O}^{2+}$ pair by removing oxygen atoms close to the interface to create V$_{\rm O}^{2+}$@T$_c$ and V$_{\rm O}^{2+}$@T$_a$, and the atomic positions of HfO$_2$ were then fully relaxed. Despite the simplicity of our model, we observed a spontaneous transformation of V$_{\rm O}^{2+}$@T$_c\rightarrow$ V$_{\rm O}^{2+}$@PO with an out-of-plane polarization while V$_{\rm O}^{2+}$@T$_a$ is much more stable. All files of optimized structures are uploaded to \url{https://github.com/sliutheorygroup/supplementary-material/tree/main/L23}. These additional calculations further corroborate our main conclusion that charged oxygen vacancies close to the metallic electrodes could induce structural polymorphism. Additionally, we report the energy difference between \Vop@Ta and \Vop@Tc between the metallic electrodes and the energy difference in bulk in Table~\ref{electrodes}, showing that the energy difference is reduced between metallic electrodes compared to that in bulk.

\begin{table}[htbp]
\centering
\caption{Comparison of energy difference between ${\rm V}_{\rm O}^{2+}$@Ta and {${\rm V}_{\rm O}^{2+}$@Tc} between metallic electrodes and in bulk. Energy in eV. 
\label{electrodes}}
\begin{tabular}{p{4cm}|p{2cm}|p{2cm}|p{3cm}}
\hline
\hline
\multirow{2}*{} &\multicolumn{3}{c}{$E$}\\
\cline{2-4}
&\Vop@T$_a$&\Vop@T$_c$&\Vop@T$_c-$\Vop@T$_a$\\
\cline{1-4}
\multirow{1}*{bulk}&$975.887$&$-975.369$&$0.518$\\
\cline{1-4}
\multirow{1}*{between electrodes}&$-1363.1506$&$-1362.8164$&$0.334$\\
\hline
\hline
\end{tabular}
\end{table}

\clearpage
\newpage

\section{results obtained using SCAN}

It is indeed the case that PBE often strongly overestimates the structural distortions for perovskite ferroelectrics. Specifically, PBE substantially overestimates the tetragonality ($\xi=c/a$) of PbTiO$_3$, predicting a value of 1.20 that is much larger than the experimental value of 1.06. However, many previous DFT investigations (including our work) have shown that PBE can accurately predict the lattice constants of different polymorphs of HfO$_2$. We performed a benchmark study using different exchange-correlation (XC) functionals including the more advanced functional such as SCAN. As reported in Table~\ref{latticeSCAN}, the lattice constants predicted with PBE are comparable with SCAN results and agree with experimental values very well (error $<0.8\%$).

Moreover, as presented in Table~\ref{defectenergySCAN}, PBE and SCAN predict nearly identical energy difference between T$_a$ and T$_c$ containing oxygen vacancy pair of different charge states: T$_a$ and T$_c$ containing charge neutral oxygen vacancies, denoted as ${\rm V}_{\rm O}^{0}$@T$_a$ and ${\rm V}_{\rm O}^{0}$@T$_c$, have comparable stability (energy difference within 0.05~eV); T$_a$ and T$_c$ containing charged oxygen vacancies, denoted as ${\rm V}_{\rm O}^{2+}$@T$_a$ and ${\rm V}_{\rm O}^{2+}$@T$_c$, have drastically different stabilities with ${\rm V}_{\rm O}^{2+}$@T$_a$ being much lower in energy than ${\rm V}_{\rm O}^{2+}$@T$_c$ by at least 0.5~eV.   

\begin{figure}[htp]
\centering
\includegraphics[width=5in]{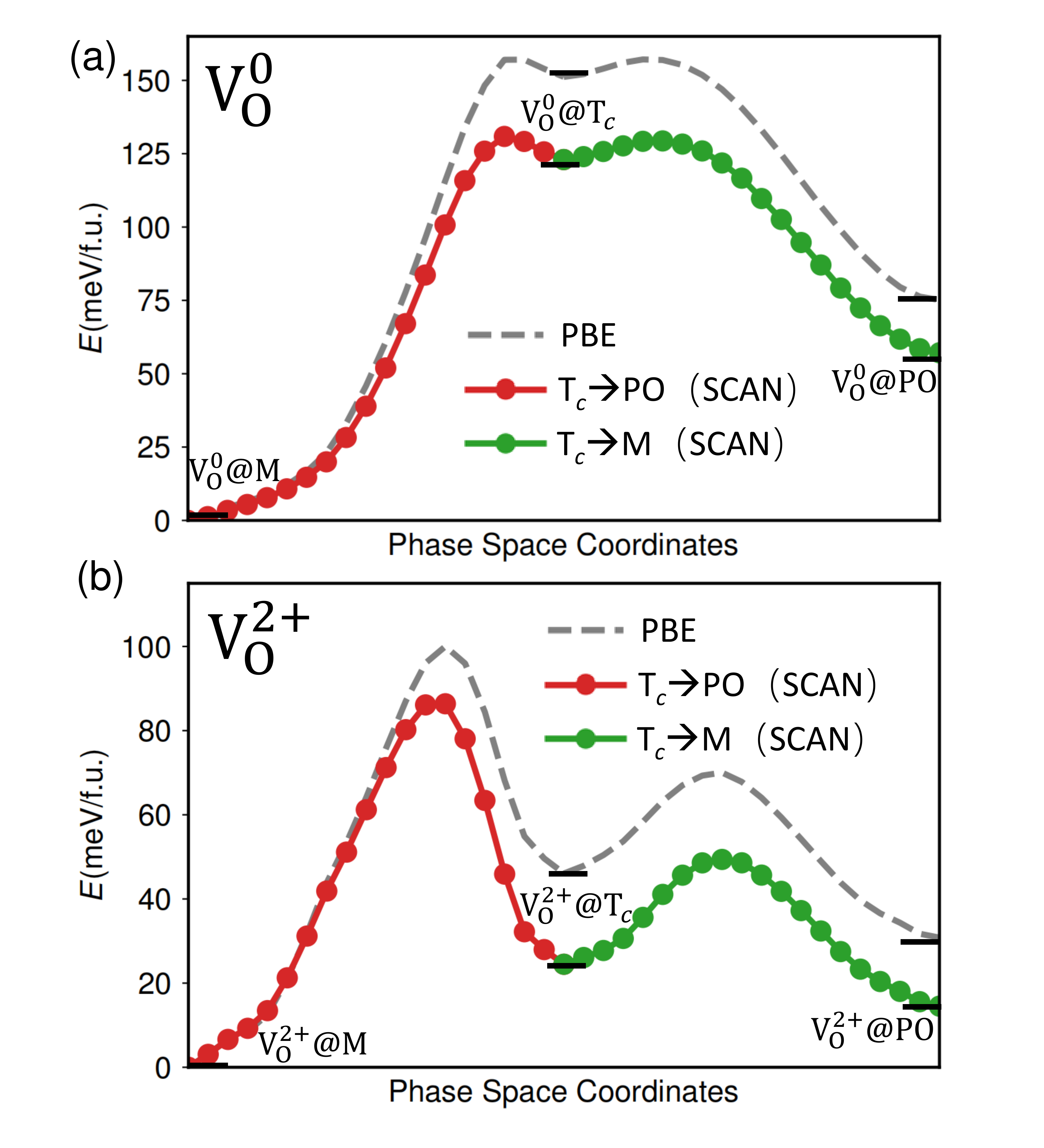}
 \caption{Comparison of minimum energy pathways identified with PBE and SCAN for phase transitions between M, T, and PO phases in the presence of (a) V$_{\rm O}^0$ and (b) V$_{\rm O}^{2+}$ at a concentration of 3.125~\%. The energy of the M phase is chosen as the zero energy point. 
}
 \label{MEP}
 \end{figure}

Finally, we also identified the minimum energy pathways (MEPs) for several representative phase transitions using SCAN with the values of energy barriers reported in Table~\ref{MEPenergy}. Figure.~\ref{MEP} compares the MEPs determined with PBE and SCAN. It is evident that PBE and SCAN predict the same energetic ordering for M, PO, and T phases and similar magnitudes of energy barriers for phase transitions. In particular, the suppression of T$_c\rightarrow$M in the presence of V$_{\rm O}^{2+}$ is more pronounced in SCAN. These additional calculations demonstrate that the key mechanisms proposed in this work are quite robust.

 \begin{table}[htp]
 	\caption{Lattice parameters in \AA~for different phases of HfO$_2$ computed with different types of XC functionals. Error relative to the experimental value\cite{MacLaren09pG103,Hoffmann15p072006,Wu20p101482} reported in the square bracket. LDA results taken from~\cite{Liu19p054404}.  \label{latticeSCAN}}
 	\begin{center}
 		\begin{tabular}{p{1cm}p{2.7cm}p{2.5cm}p{2.5cm}p{2.5cm}}
 			\hline
    \hline
    \multicolumn{2}{l}{Phase}& $a$ & $b$ & $c$\\
\hline
 		
 			\hline
    \multirow{4}*{T}&PBE(our work)& 5.225 [0.48\%] & 5.074 [0.28\%] & 5.074 [0.28\%]\\
\cline{2-5}
&SCAN&5.179 [0.40\%]&5.015 [0.89\%]&5.015 [0.89\%]\\
\cline{2-5}
&LDA&5.08 [2.3\%]&5.00 [1.19\%]&5.00 [1.19\%]\\
\cline{2-5}
&Exp.&5.20&5.06&5.06\\
\hline
    \multirow{3}*{PO}&PBE(our work)& 5.266 [0.69\%]   & 5.046 [0.72\%]  & 5.077 [0.34\%]\\
\cline{2-5}
&SCAN&5.208 [0.42\%]&4.979 [0.62\%]&5.017 [0.85\%]\\
\cline{2-5}
&LDA&5.18  [0.96\%]&4.98  [0.60\%]&5.00 
 [1.19\%]\\
\cline{2-5}
&Exp.&5.229&4.994&5.059\\
\hline
    \multirow{3}*{M}&PBE(our work)& 5.326 [0.19\%] & 5.137 [0.25\%]  & 5.188 [0.1\%]\\
\cline{2-5}
&SCAN&5.264 [0.98\%]&5.000 [2.9\%]&5.134 [0.95\%]\\
\cline{2-5}
&LDA&5.22 [1.81\%]&5.05 [1.94\%]&5.14 [0.83\%]\\
\cline{2-5}
&Exp.&5.316&5.150&5.183\\
\hline	
 			\hline
 		\end{tabular}
 	\end{center}
 \end{table}

\begin{table}[htp]
\centering
\caption{Energies of 2$\times$2$\times$2 supercells containing oxygen vacancy pair in the T phase. Energy in eV. \label{defectenergySCAN}}
\begin{tabular}{p{3cm}|p{3cm}|p{2cm}p{2cm}p{2cm}}
\hline
 
\multirow{2}*{}& \multirow{2}*{Charge state} &\multicolumn{3}{c}{$E$}\\
\cline{3-5}
&&T$_a$&T$_c$&T$_c-$T$_a$\\
\cline{1-5}
\multirow{2}*{PBE}&V$_{\rm O}^0$&$-949.282$&$-949.334$&$-0.0519$\\
\cline{2-5}
&V$_{\rm O}^{2+}$&$-975.887$&$-975.369$&0.518\\
\hline
\multirow{2}*{SCAN}&V$_{\rm O}^0$&$-2748.1654$&$-2748.1964$&$-0.031$\\
\cline{2-5}
&V$_{\rm O}^{2+}$&$-2376.2376$&$-2775.6391$&0.598\\
\hline
\end{tabular}
\end{table}

\begin{table}[htp]
\centering
\caption{Phase transition barriers ($\Delta E$) in meV per formula unit (meV/f.u.). \label{MEPenergy}}

\begin{tabular}{p{1.5cm}p{2.5cm}|p{2cm}|p{2cm}}
\hline
\multicolumn{2}{c|}{ \multirow{2}*{Phase transition} }& \multicolumn{2}{c}{$\Delta E$}\\
\cline{3-4}
\multicolumn{2}{c|}{}&{PBE}&{SCAN}\\
\hline
\multirow{2}*{V$_{\rm O}^0$}&T$_c$$\rightarrow$M$_c$-4&5.9&7.9\\
\cline{2-4}
&T$_c$$\rightarrow$PO$_c$-3&5.9&6.4\\
\hline
\multirow{2}*{V$_{\rm O}^{2+}$}&T$_c$$\rightarrow$M$_c$-4&53.9&61.8\\
\cline{2-4}
&T$_c$$\rightarrow$PO$_c$-3&24&24.9\\
\hline

\end{tabular}
\end{table}

\clearpage
\newpage

\bibliography{SL.bib}